\documentclass[sigconf, table]{acmart}

\acmConference[ICSE 2024]{46th International Conference on Software Engineering}{April 2024}{Lisbon, Portugal}

\usepackage{enumitem}

\usepackage{soul}
\usepackage{url}
\usepackage{amsthm, newtxmath}
\usepackage{balance}
\usepackage{bm}
\usepackage{booktabs}
\usepackage{color,xcolor}
\usepackage{mathtools}
\usepackage{multirow}
\usepackage{tcolorbox}
\usepackage{listings}
\usepackage{courier}
\usepackage{cancel}
\usepackage[ruled,linesnumbered]{algorithm2e}
\allowdisplaybreaks

\SetCommentSty{mycommfont}

\usepackage{pifont}

\usepackage{tcolorbox}

\newif\ifshowcomments
\showcommentstrue

\ifshowcomments
\newcommand{\yao}[1]{\mytodored{[yao: #1]}}
\else
\newcommand{\yao}[1]{}
\fi

\ifshowcomments
\newcommand{\sw}[1]{\mytodocyan{[wang: #1]}}
\else
\newcommand{\sw}[1]{}
\fi

\ifshowcomments
\newcommand{\ma}[1]{\mytodoorange{[ma: #1]}}
\else
\newcommand{\ma}[1]{}
\fi

\ifshowcomments
\newcommand{\jzl}[1]{\mytodopink{[ji: #1]}}
\else
\newcommand{\jzl}[1]{}
\fi

\newenvironment{proofskt}{%
  \proof}{\endproof}

\newcommand{\mytodored}[1]{\textcolor{red}{\ding{46}~{\sf}~#1}}

\newcommand{\mytodoorange}[1]{\textcolor{orange}{\ding{46}~{\sf}~#1}}
\newcommand{\mytodocyan}[1]{\textcolor{cyan}{\ding{46}~{\sf}~#1}}
\newcommand{\mytodopink}[1]{\textcolor{purple}{\ding{46}~{\sf}~#1}}

\newtheorem{example}{Example}[section]
\newtheorem{definition}{Definition}[section]
\newtheorem{theorem}{Theorem}[section]

\newcommand{\revise}[1]{{\color{revisecolor}{#1}}} 
\newcommand{\sm}{Supp.~Material\xspace}
\newcommand{\parh}[1]{\noindent\textbf{#1}}

\newcommand{\tool}{\textsc{CICheck}\xspace}
\newcommand{\edsan}{\textsc{ED-Check}\xspace}
\newcommand{\psan}{\textsc{P-Check}\xspace}

\newcommand{\F}{Fig.}
\newcommand{\E}{Eqn.}
\newcommand{\T}{Table}
\renewcommand{\S}{Sec.}
\newcommand{\A}{Alg.}
\newcommand{\D}{Def.}
\newcommand{\Ex}{Example}

\newcommand{\ci}{\Perp}
\newcommand{\notci}{\not\Perp}
\newcommand{\dsep}{\Perp_G}

\newcommand*\BitAnd{\mathbin{\&}}
\newcommand*\BitOr{\mathbin{|}}

\definecolor{pptred}{RGB}{176,36,24}
\definecolor{pptpurple}{RGB}{56,87,35}

\definecolor{revisecolor}{RGB}{0,0,0}

\begin{document}

\title{\revise{Enabling Runtime Verification of Causal Discovery Algorithms with Automated
Conditional Independence Reasoning\\(Extended Version)}}

\author{Pingchuan Ma}
\email{pmaab@cse.ust.hk}
\affiliation{%
  \institution{Hong Kong University of Science and Technology}
  \country{Hong Kong SAR}
}

\author{Zhenlan Ji}
\email{zjiae@cse.ust.hk}
\affiliation{%
  \institution{Hong Kong University of Science and Technology}
  \country{Hong Kong SAR}
}

\author{Peisen Yao}
\email{pyaoaa@zju.edu.cn}
\affiliation{%
  \institution{Zhejiang University}
  \country{China}
}

\author{Shuai Wang}
\authornote{Corresponding author.}
\email{shuaiw@cse.ust.hk}
\affiliation{%
  \institution{Hong Kong University of Science and Technology}
  \country{Hong Kong SAR}
}

\author{Kui Ren}
\email{kuiren@zju.edu.cn}
\affiliation{%
  \institution{Zhejiang University}
  \country{China}
}

\begin{abstract}

Causal discovery is a powerful technique for identifying causal
relationships among variables in data. It has been widely used in various
applications in software engineering. Causal discovery extensively involves
conditional independence (CI) tests. Hence, its output quality highly depends on
the performance of CI tests, which can often be unreliable in practice.
Moreover, privacy concerns arise when excessive CI tests are performed. 

Despite the distinct nature between unreliable and excessive CI tests, this
paper identifies a unified and principled approach to addressing both of them.
Generally, CI statements, the outputs of CI tests, adhere to Pearl's axioms,
which are a set of well-established integrity constraints on conditional
independence. Hence, we can either detect erroneous CI statements if they
violate Pearl's axioms or prune excessive CI statements if they are logically
entailed by Pearl's axioms. Holistically, both problems boil down to reasoning
about the consistency of CI statements under Pearl's axioms (referred to as CIR
problem).

We propose a runtime verification tool called \tool, designed to harden causal
discovery algorithms from reliability and privacy perspectives. \tool\ employs a
sound and decidable encoding scheme that translates CIR into SMT problems. To
solve the CIR problem efficiently, \tool\ introduces a four-stage decision
procedure with three lightweight optimizations that actively prove or refute
consistency, and only resort to costly SMT-based reasoning when necessary. Based
on the decision procedure to CIR, \tool\ includes two variants: \edsan\ and
\psan, which detect erroneous CI tests (to enhance reliability) and prune
excessive CI tests (to enhance privacy), respectively. We evaluate \tool\ on
four real-world datasets and 100 CIR instances, showing its effectiveness in
detecting erroneous CI tests and reducing excessive CI tests while retaining
practical performance.

\end{abstract}

\begin{CCSXML}
<ccs2012>
  <concept>
      <concept_id>10002950.10003648.10003649.10003650</concept_id>
      <concept_desc>Mathematics of computing~Bayesian networks</concept_desc>
      <concept_significance>500</concept_significance>
      </concept>
  <concept>
      <concept_id>10002978.10002986.10002990</concept_id>
      <concept_desc>Security and privacy~Logic and verification</concept_desc>
      <concept_significance>500</concept_significance>
      </concept>
  <concept>
      <concept_id>10011007.10011074.10011099</concept_id>
      <concept_desc>Software and its engineering~Software verification and validation</concept_desc>
      <concept_significance>300</concept_significance>
      </concept>
</ccs2012>
\end{CCSXML}

\ccsdesc[500]{Mathematics of computing~Bayesian networks}
\ccsdesc[500]{Security and privacy~Logic and verification}
\ccsdesc[300]{Software and its engineering~Software verification and validation}
%%
% Keywords. The author(s) should pick words that accurately describe
% the work being presented. Separate the keywords with commas.
\keywords{causal discovery, conditional independence, SMT}

\maketitle

\section{Introduction}\label{sec:intro}

Causality analysis has become an increasingly popular methodology for solving a
variety of software engineering problems, such as software
debugging~\cite{attariyan2008using,xu2015systems,fariha2020causality,
dubslaff2022causality}, performance
diagnosis~\cite{chen2014causeinfer,yoon2016dbsherlock}, DNN testing, repairing,
and explanations~\cite{zhang2022adaptive,sun2022causality,ji2023cc}. It has also
been successfully adopted in many other important domains, such as
medicine~\cite{pinna2010knockouts}, economics~\cite{addo2021exploring}, and
earth science~\cite{runge2019inferring}, and data
analytics~\cite{ma2022xinsight}.

Causal discovery is the core technique of causality analysis. It aims to
identify causal relations among variables in data and establish a causal graph.
As a result, its output constitutes the foundation of the entire causality
analysis pipeline and downstream applications. Many causal discovery algorithms,
such as PC~\cite{spirtes2000causation} and FCI~\cite{zhang2008completeness},
perform a series of conditional independence (CI) tests on the data and build
the causal graph using rules.\footnote{A CI is by $X\ci Y\mid \bm{Z}$, denoting
that $X$ is conditionally independent of $Y$ given a set oxf conditional
variables $\bm{Z}$. A CI test is a hypothesis test that examines this property.}
Thus, the output quality of causal discovery algorithms is highly dependent on
the way CI tests are performed. 
% The responsibility of CI tests in the entire causal discovery
% pipeline may vary depending on specific usage scenarios. \jzl{Remove this sentence?}

\begin{figure*}[t]
	\centering
	\includegraphics[width=0.78\linewidth]{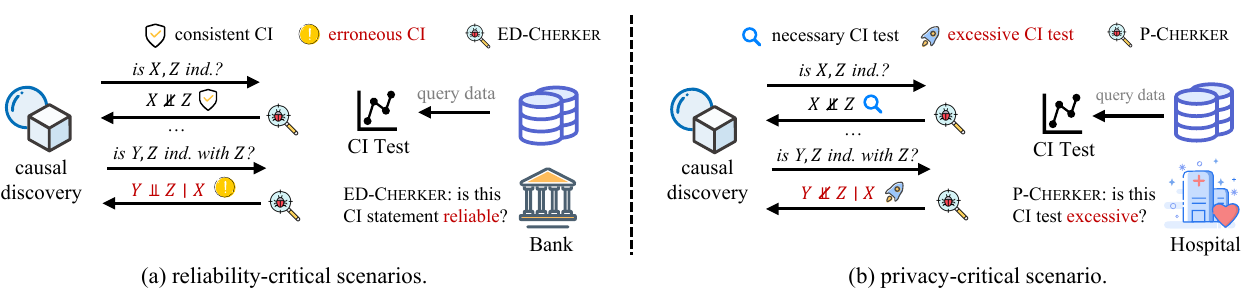}
	\vspace{-10pt}
	\caption{Two representative usage scenarios of causal discovery. ``Ind.'' denotes independence.}
	\label{fig:scenario}
	\vspace{-10pt}
\end{figure*}

\parh{Reliability and Privacy.}~\revise{This research aims to enhance two key
causal discovery scenarios: \ding{192} \textit{reliability-critical} and
\ding{193} \textit{privacy-critical}. In the \ding{192}
\textit{reliability-critical} scenario (e.g., financial data processing), the
focus is on the reliability of the causal discovery outputs. However, limited
sample size or unsuitable statistical tests may impair CI test reliability.
Unreliable CI tests may yield incorrect causal graphs, which undermines the
credibility of causal discovery. To improve the reliability, we aim to ensure
that all CI tests are consistent, a prerequisite for error-free causal
discovery. In \ding{193} \textit{privacy-critical} scenarios (e.g., health data
processing), data privacy is typically a prime concern. In accordance to common
privacy practices, access to data should be restricted unless necessary for
causal discovery. Excessive CI tests may incur unnecessary data exposure and
lead to a privacy concern. However, neither scenario has been thoroughly studied
or addressed in the literature.}

\parh{Conceptual Novelty.}~Our key insight is that we can address reliability
and privacy concerns from CI tests in a unified way, by \textit{reasoning about
the consistency of CI statements} under the integrity constraint. We refer to
this important task as Conditional Independence Reasoning (CIR). It's important
to note that CI statements within the same distribution comply with the
well-established Pearl's axioms, underpinned by statistical
theories~\cite{pearl2022graphoids}. From this, we derive two key runtime
verification strategies:

\begin{itemize}[leftmargin=*,labelindent=0pt]
	\item In \textit{reliability-critical} scenarios, we can detect erroneous
	outcomes of CI tests by checking whether the CI statements, derived from CI
	tests, comply with Pearl's axioms. If not, we conclude that the CI
	statements contain errors and may need to abort the execution or alert
	users. \textit{Example: the \textcolor{pptred}{red CI statement} is
	erroneous in \F~\ref{fig:scenario}(a) because its negation is entailed by
	existing CI statements.}
	\item In \textit{privacy-critical} scenarios, we can prune excessive CI
	tests if their outcomes can be logically entailed by prior CI statements
	(under Pearl's axioms). \textit{Example: the \textcolor{pptred}{red CI
	statement} is confirmed without incurring data access in
	\F~\ref{fig:scenario}(b), because it can be entailed by prior CI
	statements.}
\end{itemize}

\parh{Technical Pipeline.}~It is evident that the above validation schemes can
be integrated into existing causal discovery algorithms during runtime. Drawing
upon the successful application of Runtime Verification (RV) tools in software
security (e.g., preventing memory errors)~\cite{chen2016parametric,
rocsu2009runtime}, we design a novel runtime verification tool, namely \tool, to
harden the reliability and privacy of causal discovery algorithms. At the heart
of \tool is an SMT-based reasoning system aligned with Pearl's axioms, which
facilitates consistency checking in CI statements. \tool comprises two variants,
\edsan and \psan, each employing different approaches to enhance common causal
discovery algorithms in accordance with \textit{reliability-critical} and
\textit{privacy-critical} usage scenarios, respectively. 

However, efficiently implementing \tool\ is challenging. First, CIR involves
reasoning about first-order constraints, and while there are many off-the-shelf
constraint solvers, adapting them to solve CIR problems is difficult. The
challenge lies in encoding the problems soundly and compactly, while ensuring
the encoded formulas are decidable. Second, the scope of all CI statements is
significantly large, reaching up to $O(3^n)$ for problems involving $n$
variables. This exponential increase in CI statements contributes to
considerable overhead for SMT solving. To address the first challenge, we
introduce a sound and decidable encoding scheme that translates CIR into SMT
problems. To address the second challenge, we design a four-stage decision
procedure as a backend for \tool\ to solve CIR problems. Crucially, we design
three lightweight optimizations that actively prove or refute consistency, and
only resort to the costly SMT-based reasoning for instances when these
strategies don't yield a definitive resolution.

\revise{We evaluated the \edsan\ and \psan\ variants of \tool\ on four
real-world datasets and three synthetic datasets encompassing thousands of CI
tests.} The result shows the effectiveness of \edsan\ in detecting erroneous CI
tests and \psan\ in pruning excessive CI tests and reducing privacy breaches
significantly. We also evaluate 100 synthetic instances of the CIR problem and
observe that our proposed optimization schemes make the expensive, nearly
intractable SMT-based reasoning in \tool\ feasible in practice. Our
optimizations eliminate over 90\% of SMT solving procedures and substantially
cut down the execution time. In summary, we present the following contributions:

\begin{itemize}[leftmargin=*,labelindent=5pt]
	\item Given the prosperous success of causal discovery algorithms, we
	advocate validating and augmenting causal discovery algorithms with runtime
	verification techniques in terms of both reliability and privacy.
	\revise{This contribution effectively broadenes the methodological horizon
	of formal methods while addressing emerging needs in causal inference.}
	
	\item Our proposed solution, \tool, is rigorously established based on
	Pearl's axioms. It offers a sound, complete, and decidable runtime
	verification framework for causal discovery algorithms. To improve 
	performance, we present a four-stage decision procedure including SMT-based
	solving and three optimizations.
	
	\item We evaluate \tool\ on four real-world datasets and 100 synthetic
	instances of the CIR problem. The results show that \tool\ can effectively
	detect erroneous CI tests and reduce excessive CI tests. Moreover, the
	synergistic use of our optimizations makes solving the CIR problem highly
	practical. \tool\ is publicly available at an anonymous
	repository~\cite{artifact}.
\end{itemize}

\parh{Potential Impact.}~We have disseminated our findings with multiple
researchers from the causality community and received positive feedbacks. To
quote one of their remarks (the developer of causal-learn~\cite{causallearn}),
``\textit{it is indeed an awesome tool and we believe that it could definitely
benefit the community a lot.}'' This underscores our work's prospective
influence on the broader research community.

\section{Background}
\label{sec:prel}

\parh{Notations.}~We denote a random variable by a capital letter, e.g., $X$,
and denote a set of random variables by a bold capital letter, e.g., $\bm{X}$.
We use $\bm{V}$ to denote the set of all random variables. As a slight abuse of
notation, depending on the context, we also use a capital letter $X$ or a bold
capital letter $\bm{X}$ to denote a node or a set of nodes in the causal graph,
respectively. We use $\bm{X}\ci \bm{Y}\mid \bm{Z}$ to represent a general CI
relationship and use $\bm{X}\dsep \bm{Y}\mid \bm{Z}$ to represent d-separation
\revise{(i.e., a graphical separation criterion in a
graph~\cite{spirtes2000causation}; see details in \S~\ref{sm:prel} of \sm)}. A
statement of conditional independence/dependence is represented by lower Greek
letters, e.g., $\tau\coloneqq (\bm{X}\ci \bm{Y}\mid \bm{Z})$ or $\rho \coloneqq
(\bm{X}\notci \bm{Y}\mid \bm{Z})$. We also use dependent/independent CI
statements to refer to the above two classes of statements. We use $\models$ to
denote semantical entailment and $\neg$ to denote negation. For instance, if
$\tau\coloneqq (\bm{X}\ci \bm{Y}\mid \bm{Z})$, then $\neg\tau\models
(\bm{X}\notci \bm{Y}\mid \bm{Z})$. We use $\Sigma$ to denote a set of CI
statements and $\Phi$ to denote Pearl's axioms. And, we use $\textit{KB}$ to
denote a knowledge base built on $\Sigma$ and $\Phi$. $\textit{KB}\models \bot$
indicates that it is inconsistent. 
% (i.e., contains at least one conflict).

\subsection{CI \& Pearl's Axioms}
\label{subsec:ci-axiom}
\begin{definition}[Conditional Independence and Dependence]
	Let $\bm{X},\bm{Y},\bm{Z}\subseteq \bm{V}$ be three disjoint sets of random
	variables from domain $\bm{V}$. We say that $\bm{X}$ and $\bm{Y}$ are
	conditionally independent given $\bm{Z}$ (denoted by $\bm{X}\ci \bm{Y}\mid
	\bm{Z}$) if and only if $P(\bm{Z})>0$ and 
	\begin{equation}
		\label{eq:ci-def}
		\small
		P(\bm{X},\bm{Y}\mid \bm{Z}) = P(\bm{X}\mid\bm{Z})P(\bm{Y}\mid\bm{Z})
	\end{equation}
	We say that $\bm{X}$ and $\bm{Y}$ are conditionally dependent given $\bm{Z}$
	(denoted by $\bm{X}\notci \bm{Y}\mid \bm{Z}$) if and only if $P(\bm{Z})>0$ and 
	\begin{equation}
		\label{eq:cd-def}
		\small
		P(\bm{X},\bm{Y}\mid \bm{Z}) \neq P(\bm{X}\mid\bm{Z})P(\bm{Y}\mid\bm{Z})
	\end{equation}
\end{definition}

Empirically, statistical tests like $\chi^2$-squared test examine CI. Like many
statistical tests, a $p$-value indicates the error probability when rejecting
the null hypothesis. Here, the null hypothesis assumes conditional independence
between $\bm{X}$ and $\bm{Y}$ given $\bm{Z}$. Typically, the null hypothesis is
rejected if $p$-value $<1-\alpha$ ($\alpha$ is the level of significance). A
$p$-value$<1-\alpha$ rejects the null hypothesis, implying $\bm{X}\notci
\bm{Y}\mid \bm{Z}$. Otherwise, $\bm{X}\ci \bm{Y}\mid \bm{Z}$. 
% \jzl{The
% definition of $\alpha$ is missing.}
% In our context, the null hypothesis is that
% $\bm{X}$ and $\bm{Y}$ are conditionally independent given $\bm{Z}$. Usually, the
% $p$-value is compared against a threshold $1-\alpha$ to determine whether to
% reject the null hypothesis. If the $p$-value is smaller than $1-\alpha$, then
% the null hypothesis is rejected and we have $\bm{X}\notci \bm{Y}\mid \bm{Z}$;
% otherwise, we have $\bm{X}\ci \bm{Y}\mid \bm{Z}$.

Beyond CI tests on empirical distribution, efforts are made to axiomatize CI
statements based on their mathematical definition. Pearl's axioms are well-known
rules that specify the integrity constraint of CI statements within a
domain~\cite{pearl1988probabilistic}. They can be represented as \textit{Horn
clauses} over a class of probability distributions $\mathcal{P}$ to verify
them~\cite{geiger1993logical}. Here, we use $\mathcal{P}$ to denote the
universal set of probability distributions of interest. In particular, we
consider probability distributions over faithful Bayesian networks \revise{(a
standard setup in causality analysis~\cite{spirtes2000causation}). Intuitively,
it implies the correspondence between CI in the probability distribution and the
graphical structure (see the definition in \S~\ref{sm:prel} of \sm).} We
present Pearl's axioms in the form of Horn clauses as follows.

\parh{\textit{Symmetry}.}
\begin{equation}
	\small
	(\bm{X}\ci \bm{Y} \mid \bm{Z})\iff (\bm{Y}\ci \bm{X}\mid \bm{Z})
\end{equation}
\parh{\textit{Decomposition}}.
\begin{equation}
	\small
	\begin{aligned}
		(\bm{X}\ci \bm{Y}\cup \bm{W} \mid \bm{Z})&\implies 
		(\bm{X}\ci \bm{Y}\mid \bm{Z})
		% \\
		% (\bm{X}\ci \bm{Y}\cup \bm{W} \mid \bm{Z})&\implies 
		% (\bm{X}\ci \bm{W}\mid \bm{Z})
	\end{aligned}
\end{equation}
\parh{\textit{Weak Union}.}
\begin{equation}
	\small
	(\bm{X}\ci \bm{Y}\cup \bm{W}\mid \bm{Z})
	\implies 
	(\bm{X}\ci \bm{Y}\mid \bm{Z}\cup \bm{W})
\end{equation}
\parh{\textit{Contraction}.}
\begin{equation}
	\small
	(\bm{X}\ci \bm{Y} \mid \bm{Z})\land (\bm{X}\ci \bm{W}\mid Z\cup \bm{Y})
	\implies (\bm{X}\ci \bm{Y}\cup \bm{W}\mid \bm{Z})
\end{equation}
\parh{\textit{Intersection}.}
\begin{equation}
	\small
	(\bm{X}\ci \bm{Y}\mid \bm{Z}\cup \bm{W})\land (\bm{X}\ci \bm{W}\mid \bm{Z}\cup\bm{Y}) 
	\implies 
	(\bm{X}\ci \bm{Y}\cup \bm{W}\mid \bm{Z})
\end{equation}
% For faithful Bayesian networks, the following axioms also hold.

\parh{\textit{Composition}.}
\begin{equation}
	\small
	(\bm{X}\ci \bm{Y}\mid \bm{Z})\land (\bm{X}\ci \bm{W}\mid \bm{Z})
	\implies (\bm{X}\ci \bm{Y}\cup \bm{W} \mid \bm{Z})
\end{equation}
\parh{\textit{Weak Transitivity}.}
\begin{equation}
	\label{eq:wt}
	\small
	\begin{aligned}
		(\bm{X}\ci \bm{Y}\mid \bm{Z})\land (\bm{X}\ci \bm{Y}\mid \bm{Z}\cup \{U\})&\implies(U\ci \bm{Y}\mid \bm{Z})
		% \\
		% (\bm{X}\ci \bm{Y}\mid \bm{Z})\land (\bm{X}\ci \bm{Y}\mid \bm{Z}\cup \{U\})&\implies(\bm{X}\ci U\mid \bm{Z})
	\end{aligned}
\end{equation}
\parh{\textit{Chordality}.}
\begin{equation}
	\label{eq:chord}
	\small
	\begin{aligned}
		(X\ci Y \mid \{Z,W\})\land (Z\ci W \mid \{X,Y\})&\implies(X\ci Y \mid Z)
		% \\
		% (X\ci Y \mid \{Z,W\})\land (Z\ci W \mid \{X,Y\})&\implies(X\ci Y \mid W)
	\end{aligned}
\end{equation}
$X,Y,Z,W,U$ are single variables in \E~\ref{eq:wt} and \E~\ref{eq:chord}.

\parh{CI Axioms vs.~CI Tests.}~The above properties are termed \textit{axioms}
as, considering certain assumptions such as faithful Bayesian networks, there
are no probability distributions that contradict them. They essentially serve as
integrity constraints useful in refuting inaccurate CI statements.
% The above properties are called \textit{axioms}
% because under the given assumptions (e.g., faithful Bayesian networks), there
% exists no probability distribution that violates them. In other words, they are
% integrity constraints and can be used to refute erroneous CI statements.
However, they are not sufficient to deduce whether a CI statement is correct
because axioms are subject to the whole class of probability distributions while
the correctness of a CI statement is only concerned with a specific
distribution. In other words, these CI statements can only be examined by CI
tests on a particular probability distribution. In this regard, a CI test can be
deemed an oracle to CI statements. Theoretically, CI tests are asymptotically
accurate although they may be inaccurate with finite samples. Due to inaccuracy,
the axioms may be violated. The reliability- and privacy-enhancing variants of
\tool\ (\edsan\ and \psan) aim to \ding{192} check if a CI statement obtained by
CI tests complies with Pearl's axioms, and \ding{193} prune excessive CI tests
that are directly entailed by CI axioms.

\subsection{Constraint-based Causal Discovery}
\label{subsec:ccd}

In this section, we briefly introduce the application of \tool, i.e., causal
discovery.
Causal discovery aims to learn causal relations from observational data and is a
fundamental problem in causal inference. It has been widely adopted in many
fields. Causal relations among data are represented in the form of a directed
acyclic graph (DAG), where the nodes are constituted by all variables $\bm{V}$
in data and an edge $X\to Y$ in the DAG indicates a cause-effect relation
between the parent node $X$ and the child node $Y$. 

Constraint-based causal discovery is the most classical approach to identifying
the causal graph (precisely, its Markov equivalence class) from observational
data~\cite{spirtes2000causation}. It is based on the assumptions below that
affirm the equivalence between CI and d-separation. 

\begin{definition}[Global Markov Property (GMP)]
	\label{def:gmp}
	Given a distribution $P_{\bm{V}}$ and a DAG $G$, $G$ is said to be faithful
	to $P_{\bm{V}}$ if $X\dsep Y \mid Z\implies X\ci Y \mid Z$.
\end{definition}
\begin{definition}[Faithfulness]
	\label{def:faith}
	Given a distribution $P_{\bm{V}}$ and a DAG $G$, $P_{\bm{V}}$ is said to
	satisfy GMP relative to $G$ if $X\ci Y \mid Z\implies X\dsep Y \mid Z$.
\end{definition}

\parh{Peter-Clark (PC) Algorithm.}~\revise{To practically identify the causal
graph under the above assumptions, the PC algorithm is the most widely used. The
\textit{PC algorithm}, a prominent causal discovery
approach~\cite{spirtes2000causation}, consists of two principal stages: (1)
\textbf{skeleton identification} and (2) \textbf{edge orientation}. In the first
stage, the algorithm starts from a fully connected graph, iteratively searches
for separating sets ($\bm{Z}$) that d-separate nodes, consequently removing
associated edges. To expedite this process, a greedy strategy is implemented. In
the second stage, edge orientations are determined according to specific
criteria (see the full algorithm in \S~\ref{sm:pc} of  \sm). When
\D~\ref{def:gmp} and \D~\ref{def:faith} hold and CI tests are correct, the PC
algorithm is guaranteed to identify a genuine causal graph. From a holistic
view, one can comprehend the PC algorithm as iteratively executing a series of
CI tests, which aims to construct a causal graph with pre-defined rules and
results of CI tests.} 

\section{Research Overview}
\label{sec:overview}
In this section, we present an illustrative example to show how \tool\ can be
reduced to CIR problem solving. Then, we formulate the CIR problem and show how
to connect it with \tool.

\subsection{Illustrative Example}
\label{subsec:example}

We present a case in accordance with \F~\ref{fig:scenario} where \edsan\ detects
an erroneous CI statement in the PC algorithm. To ease presentation, we analyze
the case from the view of \edsan\ while the procedure also applies to \psan.

\begin{figure}[!htbp]
	\vspace{-5pt}
	\centering
	\includegraphics[width=0.88\columnwidth]{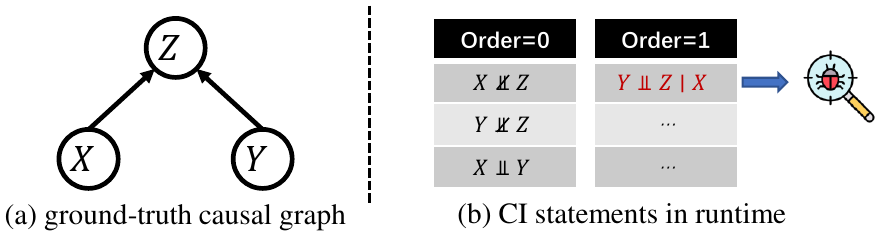}
	\vspace{-10pt}
	\caption{An erroneous CI statement occurred when launching the PC algorithm.}
	\vspace{-10pt}
	\label{fig:example}
\end{figure}

Consider a set of random variables $\bm{V}=\{X,Y,Z\}$ whose ground-truth causal
graph is shown in \F~\ref{fig:example}(a). The PC algorithm aims to identify the
causal structure from the dataset corresponding to $\bm{V}=\{X,Y,Z\}$. When
$\textit{order}=0$, the algorithm enumerates all (pairwise) marginal
independence relations.\footnote{\textit{order} is a parameter in the algorithm,
denoting the condition size of current CI tests.} As shown in the left-side
table in \F~\ref{fig:example}(b), it identifies that $\{ (X\notci Z), (Y\notci
Z), (X\ci Y)\}$. Since $Y\notci Z$, when $\textit{order}=1$, the PC algorithm
would further query whether $Y\ci Z\mid X$ holds by invoking a CI test. However,
due to errors occurred in the current CI test, $\sigma=Y\ci Z\mid X$ is deemed
true. Hence, the PC algorithm incorrectly concludes that $Y$ and $Z$ are
non-adjacent, and yields an erroneous causal graph accordingly. 

The above error is detectable under Pearl's axioms: $\neg\sigma$ can be logically
entailed by considering the marginal CI statements (shown in the left-side table
of \F~\ref{fig:example}(b)) and Pearl's axioms, which can be proved easily by
contradiction. Consider the proof below:
\begin{proof}
	Suppose $Y\ci Z\mid X$ is true. By Contraction axiom, $Y\ci Z\mid X$ and $X\ci
	Y$ implies that $Y\ci \{X, Z\}$. Then, by Decomposition axiom, $Y\ci \{X, Z\}$
	implies that $Y\ci X$ and $Y\ci Z$. This contradicts the fact that $X\notci Y$.
\end{proof}

\vspace{-3pt}
\revise{Therefore, $\sigma$ is not warranted by Pearl's axioms and we can flag
it as inconsistent. } 

\parh{Extension to \psan.}~Similar to earlier case, \psan\ preempts CI testing
by concurrently validating if $Y\ci Z\mid X$ or $Y\notci Z\mid X$ is warranted.
Given that Pearl's axioms only warrant $Y\notci Z\mid X$, we directly infer
$Y\notci Z\mid X$ as the correct statement. \revise{Besides, inaccurate CI tests
may undermine the effectiveness of \psan. We discuss this issue in
\S~\ref{subsec:impl} and \S~\ref{sec:discussion}.}

\subsection{Problem Statement}
\label{subsec:problem}

\parh{The CIR Problem.}~The above reasoning process in the proof is ad-hoc. To
deliver an automated reasoning-based runtime verification, we need a general
reasoning tool to automate the above process. We define the following problem
(Conditional Independence Reasoning; CIR) to formalize the above reasoning
process.

\begin{tcolorbox}[title=\textsc{CIR Problem Statement}, boxsep=1pt,left=2pt,right=2pt,top=0pt,bottom=1pt]
	
	\parh{Input:} $\Sigma=\{\sigma_1,\ldots,\sigma_k\}$, where
	$\sigma_1,\ldots,\sigma_k$ are CI statements.
	
	\parh{Output:} Whether $\textit{KB}\coloneq \{\Sigma, \Phi \}\models \bot$,
	where $\Phi$ is Pearl's axioms.
	
\end{tcolorbox}

If the CIR problem returns a true value, we conclude that there is an
inconsistency error among $\Sigma$ under Pearl's axioms. Otherwise, $\Sigma$ is
consistent under Pearl's axioms.

\smallskip\parh{Building \tool\ via CIR Reasoning}~\F~\ref{fig:scenario}
illustrates two typical and representative usage scenarios of \tool, in the form
of reliability- and privacy-enhancing variants, \edsan\ and \psan, respectively.
We hereby build them, by recasting their specific objective into one (or
multiple) instances of the CIR problem. In this way, the two variants of \tool\ can
be implemented in a unified manner. \A~\ref{alg:cisan} presents the algorithmic
details of \tool.

\begin{algorithm}[!htbp]
	\footnotesize
	\caption{Two variants of \tool}
	\label{alg:cisan}
	\KwIn{Variable Pair $X, Y$, Condition Set $\bm{Z}$}
	\KwOut{$X\ci Y\mid \bm{Z}$ or $X\notci Y\mid \bm{Z}$}
	\SetAlgoNoEnd
	\SetKwFunction{ErrorDetection}{\edsan}
	\SetKwFunction{Pruning}{\psan}
	\SetKwProg{Fn}{Function}{:}{}
	\Fn{\ErrorDetection{$X, Y, \bm{Z}$}}{
		$\gamma\leftarrow \textit{CITest}(X, Y, \bm{Z})$\;
		$\texttt{assert}~\textit{KB}\coloneq \{\Sigma\cup\{\gamma\}, \Phi \}\not\models \bot$\;
		\KwRet{$\gamma$}\;
	}
	\Fn{\Pruning{$X, Y, \bm{Z}$}}{
		$\gamma\leftarrow X\ci Y\mid \bm{Z}$\;
		\uIf{$\textit{KB}\coloneq \{\Sigma\cup\{\gamma\}, \Phi \}\models \bot$}{
			\KwRet{$X\notci Y\mid \bm{Z}$}
		}
		\uElseIf{$\textit{KB}\coloneq \{\Sigma\cup\{\neg\gamma\}, \Phi \}\models \bot$}{
			\KwRet{$X\ci Y\mid \bm{Z}$}}
		\Else{\KwRet{$\textit{CITest}(X, Y, \bm{Z})$}}
	}
\end{algorithm}

\textit{\edsan} initiates the CI test (line 2), obtaining the output CI
statement $\gamma$. It then checks if $\gamma$ is consistent under Pearl's
axioms (line 3), ensuring that $\textit{KB}\coloneq {\Sigma\cup{\gamma}, \Phi
}\not\models \bot$. If $\textit{KB}\models\bot$, \edsan\ can either abort
execution or notify the user. In this way, \edsan\ can detect erroneous
CI statements.

\textit{\psan} checks the consistency of both $X\ci Y\mid \bm{Z}$ and $X\notci
Y\mid \bm{Z}$ under Pearl's axioms. \revise{In particular, before conducting the
CI test (line 12; which accesses the private user data), \psan first checks
whether the incoming CI statement can be deduced from the existing CI statements
(lines 7 and 9). If it is deducible, we omit the CI test and directly return the
deduced CI statement (lines 8 and 10). Otherwise, we resort to the CI test (line
12). Hence, there is at most one CI test performed throughout the process. 

Notably, \psan\ reduces privacy leakage by pruning unnecessary CI tests. This is
because eliminating a CI test prevents access to its associated private data.
For instance, in a hospital scenario, we may deduce one CI relation among
multiple diseases (e.g., $D_1\notci D_2\mid D_3$) from the existing CI
statements (e.g., $D_1\notci D_2$, $D_1\ci D_3$ and $D_2\notci D_3$). This
eliminates the need to access the private information of $D_1\notci D_2\mid
D_3$.}% \sw{an example like hospital scenario?}

%\smallskip
\parh{Challenges.}~While the CIR problem appears simple, solving it efficiently
is far from trivial. On a theoretical level, it requires a sound, compact, and
decidable encoding scheme to convert the CIR problem into an SMT problem.
Practically speaking, the states of CI statements can be overwhelming, resulting
in as many as $O(3^n)$ CI statements for problems involving $n$ variables. This
exponential growth can result in significant overhead for SMT solving and make
solving complex CIR problems impractical. Next, we show how these technical
challenges are addressed in \S~\ref{sec:design}.
\section{Automated Reasoning for CIR}
\label{sec:design}

This section presents an automated reasoning framework for solving any instances
of the CIR problem (defined in \S~\ref{sec:overview}). We first show how CIR can
be formulated as an SMT problem, and present a decidable, SMT-based encoding
scheme to the problem. We then present several optimizations to boost the
solving of this problem, which makes it practical in real-world usage.

\subsection{Encoding the CIR Problem}
\label{subsec:basic}

\parh{Formulation.}~Our research leverages the observation that the CIR problem,
based on Pearl's axioms, can translate to a satisfiability issue. Consider an
instance of the CIR problem; we expect the solver to verify whether
$\langle\Sigma,\Phi\rangle\coloneqq \bigwedge\limits_{i=1}^k \sigma_i \wedge \bigwedge\limits_{j=1}^m \Phi_j$
is satisfiable, where $\sigma_i$ is a CI statement and $\Phi_j$ is a rule of
Pearl's axioms. If the formula is unsatisfiable, we can conclude that the CIR
instance yields a contradiction (i.e., $\textit{KB}\models\bot$).

\begin{example}
	\label{ex:1}
	Consider the example in \F~\ref{fig:example}. The CIR problem is to check
	whether the following formula is satisfiable:
	\begin{equation}
		\small
		\{X\notci Z\} \wedge \{Y\notci Z\} \wedge \{X\ci Y\} \wedge \{Y\ci Z\mid X\} \wedge \bigwedge\limits_{j=1}^m \Phi_j 
	\end{equation}
	where $\Phi_j$ represents a rule of Pearl's axioms. If the formula is
	unsatisfiable, then  the CI statements are inconsistent under Pearl's
	axioms.
\end{example}

Despite the simplicity of the procedure above, it is non-trivial to
automatically reason about the satisfiability of $\langle\Sigma,\Phi\rangle$ in
a sound, decidable, and efficient manner. First, the space of all possible CI
statements should be appropriately modeled, where unseen CI statements are also
considered. \revise{It is noteworthy that, while addressing the consistency of
CI statements, some (intermediate) CI statements do not necessarily appear in
the set. We present an example in \S~\ref{subsec:example} to illustrate this.
The check process first deduces an unseen CI statement (i.e., $Y\ci {X,Z}$).
Then, it uses this previously unseen CI statement to refute the consistency.
Consequently, simply ignoring unseen CI statements may lead to an incorrect
result.} Second, since CI statements are defined on three sets of variables
(i.e., $\bm{X},\bm{Y},\bm{Z}$), an efficient encoding of sets and their
primitives is necessary. Third, we need to encode Pearl's axioms as quantified
first-order formulas. We address them in what follows.

\smallskip
\parh{Modeling CI Statements.}~We first show how to model all possible CI
statements as SMT formulas, including potentially unseen CI statements in
$\Sigma$. We utilize the theory of Equality with Uninterpreted Functions
(EUF)~\cite{ackermann1954solvable,DBLP:series/txtcs/KroeningS16} for this
modeling. EUF extends the Boolean logic with equality predicates in the form of
$\texttt{UF}(x)=y$, where \texttt{UF} is an uninterpreted function symbol.
Notably, $\texttt{UF}$ adheres to the \textit{axiom of functional congruence},
implying identical inputs yield identical outputs (for instance, $x=y\land
\texttt{UF}(x)\neq \texttt{UF}(y)$ is unsatisfiable). This inherent feature
holds importance in CI statement encoding as it guarantees their invariance for
fixed $\bm{X},\bm{Y},\bm{Z}$. Besides, modern SMT solvers for the EUF theory
have efficient enforcement for the functional congruence axiom.

Hence, the characteristics of the EUF theory make it both suitable and practical
to model CI statements. An uninterpreted function symbol
$\texttt{CI}:2^{\bm{V}}\times 2^{\bm{V}}\times 2^{\bm{V}}\to {0,1}^2$ is
defined, having three arguments $\bm{X},\bm{Y},\bm{Z}$ that are $\bm{V}$'s
subsets. $\texttt{CI}(\bm{X},\bm{Y},\bm{Z})$ outputs a size-2 bit-vector
($b_{n-1}\ldots b_0$, $b_i \in {0, 1}$ sequence). The semantics of
$\texttt{CI}(\bm{X},\bm{Y},\bm{Z})$'s outputs are defined next.
\begin{equation}
	\label{eq:ci}
	\small
	\texttt{CI}(\bm{X},\bm{Y},\bm{Z})\coloneqq\begin{cases}
		\texttt{01} & \text{if } \bm{X}\ci\bm{Y}\mid\bm{Z} \\
		\texttt{00} & \text{if } \bm{X}\notci\bm{Y}\mid\bm{Z} \\
		\texttt{11} & \text{if } \bm{X},\bm{Y},\bm{Z} \text{ are invalid}
	\end{cases}
\end{equation}
Here, we create an additional state $\texttt{CI}(\bm{X},\bm{Y},\bm{Z})=-1$ to
ensure that the uninterpreted function $\texttt{CI}$ is well-defined for all
possible $\bm{X},\bm{Y},\bm{Z}$ inputs. $\bm{X},\bm{Y},\bm{Z}$ are invalid in
the following case:
\begin{enumerate}
	\item $\bm{X}\cap\bm{Y}\neq\emptyset$; or $\bm{X}\cap\bm{Z}\neq\emptyset$;
	or $\bm{Y}\cap\bm{Z}\neq\emptyset$.
	\item $\bm{X}=\emptyset$; or $\bm{Y}=\emptyset$.
\end{enumerate}

\parh{Representing Sets.}~Nevertheless, the argument types in the definition of
$\texttt{CI}$ given above --- sets --- are not supported in many
state-of-the-art SMT solvers. The Z3 solver, a widely-used SMT solver with
limited set support, relies on arrays to simulate sets and is inefficient. To
overcome this, we use the bit-vector theory to deliver a \textit{compact and
decidable} encoding of $\bm{X},\bm{Y},\bm{Z}$, as well as set primitives and
Pearl's axioms over these variables.

We define a function
\texttt{set2vec} 
that encodes each set variable into a bit-vector variable:
	$\texttt{set2vec}: 2^{\bm{V}}  \to \{0,1\}^{|\bm{V}|}$
Next, we use $BV_{\bm{X}}, BV_{\bm{Y}},BV_{\bm{Z}}$ to represent the bit-vector
variables corresponding to  $\bm{X}, \bm{Y},\bm{Z}$, respectively. The
bit-vector variables $BV_{\bm{X}}, BV_{\bm{Y}},BV_{\bm{Z}}$ can constitute the
arguments of the uninterpreted function $\texttt{CI}$ (with a slight abuse of
notation, we ``override'' the symbol $\texttt{CI}$ (\E~\ref{eq:ci}) and make it
to take bit-vectors as input).

With the bit-vector representation of a set, we now encode set primitives using
bit-vector operations:

\begin{align}
	\label{eq:set-primitive}
	\small
	\bm{X} = \emptyset & \iff BV_{\bm{X}} = \bm{0} \\
	|\bm{X}| = 1 &\iff \sum_{i=0}^{|\bm{V}|-1} \texttt{extract}(i, i, BV_{\bm{X}})  = 1 \\
	&\text{abbreviated to } |BV_X| = 1 \\
	\bm{X}\cup \bm{Y} & \iff BV_{\bm{X}} \BitOr BV_{\bm{Y}} \\
	\bm{X}\cap \bm{Y} & \iff BV_{\bm{X}} \BitAnd BV_{\bm{Y}} \\
	\bm{X}\subseteq \bm{Y} & \iff BV_{\bm{X}} \BitAnd BV_{\bm{Y}} = BV_{\bm{X}} \\
	X \in\bm{Y} & \iff |BV_X| = 1 \land BV_{X} \BitAnd BV_{\bm{Y}} = BV_{X}
\end{align}

Next, we show an instantiation of the encoding scheme.
\begin{example}
	\label{ex:2}
	Consider $\bm{V}=\{X,Y,Z\}$, $\bm{X}=\{X\}$, $\bm{Y}=\{Y\}$, and
	$\bm{Z}=\{Y\}$.  Without loss of generality, we have $BV_{\bm{X}}=001$,
	$BV_{\bm{Y}}=010$, $BV_{\bm{Z}}=100$ and $BV_{\emptyset}=000$. Then, the CI
	statements in \Ex~\ref{ex:1} where $\Sigma \coloneqq \{ (X\notci Z),
	(Y\notci Z), (X\ci Y), (Y\ci Z\mid X)\}$ can be encoded as:
	\begin{equation}
		\small
		\begin{aligned}
			\texttt{CI}(BV_{\bm{X}}, BV_{\bm{Z}}, BV_{\emptyset}) = \texttt{CI}(\texttt{\underline{001}, \underline{100}, \underline{000}}) = \texttt{00} \\
			\texttt{CI}( BV_{\bm{Y}}, BV_{\bm{Z}}, BV_{\emptyset}) = \texttt{CI}(\texttt{\underline{010}, \underline{100}, \underline{000}}) = \texttt{00} \\
			\texttt{CI}( BV_{\bm{X}}, BV_{\bm{Y}}, BV_{\emptyset}) = \texttt{CI}(\texttt{\underline{001}, \underline{010}, \underline{000}}) = \texttt{01} \\
			\texttt{CI}( BV_{\bm{Y}}, BV_{\bm{Z}}, BV_{X}) = \texttt{CI}(\texttt{\underline{010}, \underline{100}, \underline{001}}) = \texttt{01} \\
		\end{aligned}
	\end{equation}
\end{example}

As shown in \Ex~\ref{ex:2}, the CI statements are encoded as EUF constraints
using bit-vector variables. Similarly, we can encode the other CI relations.

\smallskip
\parh{Encoding Axioms.}~We only apply Pearl's axioms to valid arguments of the
uninterpreted function \texttt{CI}. Therefore, prior to encoding Pearl's axioms,
we define the validity constraint to check whether the argument of \texttt{CI}
is valid; these validity constraints form the premises of all axioms. 
\begin{equation}
	\small
	\label{eq:validity}
	\begin{aligned}
		\texttt{Valid}&(BV_{\bm{X}}, BV_{\bm{Y}}, BV_{\bm{Z}})\coloneqq  \\
		&BV_{\bm{X}} \BitAnd BV_{\bm{Y}} = \bm{0} 
		\land BV_{\bm{X}} \BitAnd BV_{\bm{Z}} = \bm{0} \\
		&\land BV_{\bm{Y}} \BitAnd BV_{\bm{Z}} = \bm{0} \land BV_X \neq \bm{0} \land BV_{\bm{Y}} \neq \bm{0}
	\end{aligned}
\end{equation}
Next, we encode Pearl's axioms. For each rule, we create free-variables
corresponding to the variables used in this rule, and use universal
quantification to specify the rule in all possible concretization to these
variables. 

\begin{example}
	\label{ex:3}
	Consider the Decomposition axiom:
	\begin{equation*}
		\small
		\begin{aligned}
			(\bm{X}\ci \bm{Y}\cup \bm{W} \mid \bm{Z})&\implies 
			(\bm{X}\ci \bm{Y}\mid \bm{Z})\\
			(\bm{X}\ci \bm{Y}\cup \bm{W} \mid \bm{Z})&\implies 
			(\bm{X}\ci \bm{W}\mid \bm{Z})
		\end{aligned}
	\end{equation*}
	We define the following constraint to encode the axiom (and we omit the
	validity constraint for ease of presentation).
	\begin{equation}
		\small
		\begin{aligned}
			\forall &BV_{\bm{X}},BV_{\bm{Y}},BV_{\bm{Z}},BV_{\bm{W}}\in \{0,1\}^{|\bm{V}|}\\
			& \texttt{CI}(BV_{\bm{X}}, BV_{\bm{Y}} | BV_{\bm{W}}, BV_{\bm{Z}})=\texttt{01}\implies \texttt{CI}( BV_{\bm{X}}, BV_{\bm{Y}}, BV_{\bm{Z}})=\texttt{01}\\
			& \texttt{CI}( BV_{\bm{X}}, BV_{\bm{Y}} | BV_{\bm{W}}, BV_{\bm{Z}})=\texttt{01}\implies \texttt{CI}(BV_{\bm{X}}, BV_{\bm{W}}, BV_{\bm{Z}})=\texttt{01}
		\end{aligned}
	\end{equation}
	In the above constraint, we first create a universal quantifier and four
	free bit-vector variables $BV_{\bm{X}}$, $BV_{\bm{Y}}$, $BV_{\bm{Z}},
	BV_{\bm{W}}$ corresponding to $\bm{X},\bm{Y},\bm{Z},\bm{W}$ respectively.
	Then, we create two implication constraints ($\implies$) corresponding to
	the two Horn clauses of the Decomposition axiom, where the set union is
	encoded as bitwise-or operations.
\end{example}
Similarly, we can encode the remaining axioms via universal quantifiers (see the
encoding in \S~\ref{sm:enc} of \sm).

\begin{proposition}
	\label{prop:decidable}
	The encoded SMT problem is a sound and complete representation of the CIR
	problem. And, the problem is decidable.
\end{proposition}

\begin{proofskt} 
    
    \textit{Soundness \& Completeness.} Let's consider an instance of CIR
	problem with $\Sigma$ being the set of CI statements and let $\phi_c$ be an
	oracle to CIR problem. Following the encoding procedure, we can construct an
	SMT problem with $\mathcal{C}$ being the constraints and let $\phi_s$ be an
	SMT solver. The soundness of the encoding scheme implies that if
	$\phi_s(\mathcal{T})$ returns satisfiable, then $\phi_c(\Sigma)$ returns
	consistent. The completeness of the encoding scheme implies that if
	$\phi_c(\Sigma)$ returns consistent, then $\phi_s(\mathcal{T})$ returns
	satisfiable. 

    \begin{lemma}
        \label{lemma:ci-sat}
        If there exists a joint distribution $P$ that admits all CI statements
        in $\Sigma$, then $\Sigma$ is consistent.
    \end{lemma}

    \begin{lemma}
        \label{lemma:ci-unsat}
        If there does not exist a joint distribution $P$ that admits all CI
        statements in $\Sigma$, then $\Sigma$ is inconsistent.
    \end{lemma}

    \begin{lemma}
        \label{lemma:ci-neg}
        There does not exist a joint distribution $P$ that simultaneously admits
        one CI statement and its negation.
    \end{lemma}

    \begin{lemma}
        \label{lemma:pa-sat}
        Constraints only containing the Pearl's axioms are always satisfiable.
    \end{lemma}

	Lemma~\ref{lemma:ci-sat} and Lemma~\ref{lemma:ci-unsat} are direct
	consequences by the definition of Pearl's axioms. Lemma~\ref{lemma:ci-neg}
	is a direct consequence of the definition of conditional independence
	(\E~\ref{eq:ci-def} and \E~\ref{eq:cd-def}). Lemma~\ref{lemma:pa-sat} (i.e.,
	self-consistency of Pearl's axioms) can be derived from the graphical
	criterion of conditional independence~\cite{pearl2022graphoids}.

    Assume that $\phi_s(\mathcal{T})$ is satisfiable. From this premise, it
    directly follows that there exists an uninterpreted function $\texttt{CI}$
    satisfying Pearl's axioms. We define $\texttt{CI}$ as in \E~\ref{eq:ci},
    ensuring that it is well-defined for all potential inputs
    $\bm{X},\bm{Y},\bm{Z}$. Given the existence of $\texttt{CI}$, we can
    construct a corresponding direct acyclic graph (DAG) $G$ in accordance to
    the SGS algorithm~\cite{spirtes2000causation}. Then, based on the result
    of~\cite{meek1995strong}, there exists a joint distribution $P$ that admits
    $G$. By Lemma~\ref{lemma:ci-sat}, we affirm that the CI statements are
    consistent under Pearl's axioms when the encoded SMT problem is satisfiable.
    Conversely, suppose that $\phi_s(\mathcal{T})$ is unsatisfiable. By
    Lemma~\ref{lemma:pa-sat}, we know that the minimum unsatisfiable core
    contains at least one CI statement. Without loss of generality, we assume
    $$\{\sigma_{i-1},\sigma_{i},\ldots,\sigma_j, \Phi_k,\ldots,\Phi_l\}$$ be the
    minimum unsatisfiable core of $\mathcal{T}$. We have
    $$\{\sigma_{i},\ldots,\sigma_j, \Phi_k,\ldots,\Phi_l\}\models
    \neg\sigma_{i-1}$$. In the meantime, we also have $\sigma_{i-1}$ in
    $\Sigma$. By Lemma~\ref{lemma:ci-neg}, we know that there does not exist a
    joint distribution $P$ that simultaneously admits $\sigma_{i-1}$ and its
    negation. By Lemma~\ref{lemma:ci-sat}, this indicates that the CI statements
    are inconsistent under Pearl's axioms when the encoded SMT problem is
    unsatisfiable. Therefore, the encoded SMT problem gives a sound and complete
    representation of the CIR problem, which deals with the consistency of CI
    statements.
    
    \textit{Decidability.} In our encoded SMT problem, we find constraints that
    belong to the combined theory of uninterpreted functions and quantified
    bit-vectors. According to Wintersteiger et
    al.~\cite{wintersteiger2013efficiently}, it is known that this combined
    theory is decidable. This ensures the decidability of the constraints in our
    problem, hence confirming the decidability of the entire encoded SMT
    problem. 
\end{proofskt}

\subsection{Solving the CIR Problem}

We describe the encoding of the CIR problem in \S~\ref{subsec:basic}. Despite
the rigorously formed encoding, this approach is
\textit{NEXP-complete}~\cite{wintersteiger2013efficiently}, meaning that it may
become intractable for complex problem instances. In this section, we present
several optimizations and design choices to make \tool\ practical at a moderate
cost.

\begin{figure}[t]
	%\vspace{-5pt}
	\centering
	\includegraphics[width=\linewidth]{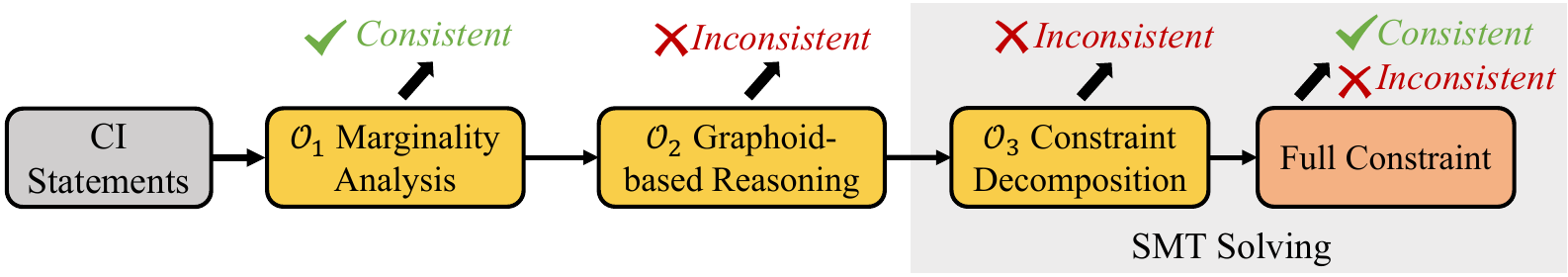}
	\vspace{-20pt}
	\caption{Overview of the CIR problem solving pipeline.}
	\vspace{-15pt}
	\label{fig:overview}
\end{figure}

\F~\ref{fig:overview} outlines the workflow of our solving procedure. In short,
before launching the SMT solver over the full constraint obtained by our
encoding scheme, we leverage the following three lightweight optimizations to
prove or refute the satisfiability of the constraint and presumably
``early-stop'' the reasoning to avoid costly SMT solving. Particularly,
Marginality Analysis is performed first and can prove the satisfiability of the
constraint. If Marginal Analysis fails, we then apply the following two
optimizations (Graphoid-based Reasoning and Constraint Decomposition) to refute
the satisfiability. SMT solving over the full constraint is only performed when
none of the optimizations reaches a conclusive result.

\subsubsection{$\mathcal{O}_1$ Marginality Analysis}
\label{subsec:ma}
In general, we need to check every CI test to ensure that each one is valid (or
can be pruned), depending on the application. However, we notice that when
$\Sigma$ only contains non-degenerate marginal CI statements, check is
\textit{not} necessary. The following theorem characterizes this property:

\begin{theorem}
	\label{thm:non-degenerate}
	Let $\Sigma$ be a set of non-degenerate marginal CI statements. Here, $\Sigma$
	is said to be non-degenerate if $\Sigma$ does not contain both a marginal CI
	statement and its negation. A marginal CI statement is a CI statement with an
	empty condition set, such as $X\ci Y$. Then, $\Sigma$ is always consistent under Pearl's axioms.
\end{theorem}

\begin{proofskt}
	By definition, $\Sigma$ is consistent under Pearl's axioms if
	there exists a joint distribution $P$ such that $P$ admits every CI
	statement in $\Sigma$. Below, we demonstrate that such a joint distribution
	$P$ always exists for arbitrary $\Sigma$, as long as it only contains
	non-degenerate marginal CI statements. Let $G_{u}$ be an undirected graph
	where each node in $G_{u}$ represents a random variable with respect to $P$.
	Two nodes $X,Y$ are connected by an edge if and only if $X\notci Y$.
	According to the ``acyclic orientation'' in~\cite{stanley1973acyclic}, any
	undirected graph can be converted to a directed acyclic graph (DAG). Let
	$G_d$ be a DAG of $G_u$. Then, according to Theorem 5
	of~\cite{meek1995strong}, a faithful multinomial $P$ exists for any DAG
	$G_{d}$.\footnote{Precisely, Theorem 5 in~\cite{meek1995strong} implies that
		there is almost always a $P$ that is faithful to $G_d$ in a
		measure-theoretic sense. The Lebesgue measure (conceptually similar to
		probability) of the set of unfaithful distributions $P$ is zero.} In this
	regard, $\Sigma$ is at least warranted in $P$ and is thus consistent.
\end{proofskt}

Theorem~\ref{thm:non-degenerate} directly implies we can omit \tool, when
$\Sigma$ exclusively holds non-degenerate marginal CI statements. Often, this
occurs in the initial phases of common causal discovery algorithms, shown in
the example below.

\begin{example}
	Consider the first three CI statements in Example~\ref{ex:1}, with $\Sigma
	\coloneqq { (X\notci Z), (Y\notci Z), (X\ci Y)}$. As $\Sigma$ consists
	solely of non-degenerate marginal CI statements, according to
	Theorem~\ref{thm:non-degenerate}, we can directly infer that $\Sigma$ is
	consistent.
\end{example}

\subsubsection{$\mathcal{O}_2$ Graphoid-based Reasoning}

In causal inference, CI statements are often represented as
graphoids~\cite{pearl2022graphoids}, which are special graphical data structures
that enable efficient reasoning about \textit{\ding{192} whether a CI statement
is always independent}. {As a result,} graphoid-based reasoning cannot
distinguish ``\textit{\ding{193} a CI statement is always dependent}'' from
``\textit{\ding{194} a CI statement can be either independent or dependent}''.
Despite this limitation, graphoid-based reasoning is useful in our context owing
to its efficiency. In practice, we find graphoid-based reasoning to be
significantly faster than solving standard CIR problems. Below, we show the
translation from an instance of the CIR problem to a sequence of graphoid-based
reasoning.

Let the knowledge base $\Sigma=\{\sigma_1,\ldots, \sigma_k\}$ be the input of a
CIR problem. We split $\Sigma$ into two parts: (1) the first part, $\Sigma_1$,
contains all dependent CI statements (e.g., $X\notci Y\mid \bm{Z}$), and (2) the
second part, $\Sigma_2$, contains all independent CI statements (e.g., $X\ci
Y\mid \bm{Z}$). We can then use graphoid-based reasoning to check if $\Sigma_2$
implies the negation of any instances in $\Sigma_1$ (e.g., $\Sigma_2\implies
\neg\sigma_i$ where $\sigma_i\in\Sigma_1$). If this occurs, we can conclude that
$\Sigma$ is inconsistent.

\begin{example}
	Consider the CI statements in Example~\ref{ex:1}, with $\Sigma \coloneqq \{
	(X\notci Z), (Y\notci Z), (X\ci Y), (Y\ci Z\mid X)\}$. Initially, to
	evaluate $\Sigma$'s consistency, we partition it into $\Sigma_1={(X\notci
	Z), (Y\notci Z)}$ and $\Sigma_2={(X\ci Y), (Y\ci Z\mid X)}$. Then, using
	graphoid-based reasoning, we verify if $\Sigma_2$ suggests either $X\ci Z$
	or $Y\ci Z$. If $\Sigma_2$ infers any of these two statements, $\Sigma$ is
	deemed inconsistent.
\end{example}

To boost the standard SMT constraint-solving process, one may sample a few concrete
inputs to rapidly disprove the satisfiability of a
constraint~\cite{chandramohan2016bingo}. Similarly, graphoid-based reasoning
serves as a lightweight testing-based error detecting scheme. With this
optimization, we may avoid costly SMT solving without compromising the
completeness of the overall CIR solving procedure.

\subsubsection{$\mathcal{O}_3$ Constraint Decomposition}
\label{subsubsec:slice}
While the graphoid-based reasoning is efficient, it cannot sufficiently leverage
the dependent CI statements in the knowledge base. Hence, an SMT solver is still
required, leading to a significant overhead. To address this issue, we propose
applying constraint decomposition to the CIR problem using domain knowledge.
This method segments the full constraint set into smaller subsets, resolved
independently and concurrently.

The key insight is that inconsistencies among CI statements frequently arise
from two sources: \ding{192} individual axioms and \ding{193} a small, closely
related subset of CI statements. Technically, given an incoming CI statement
$\gamma$, we check whether it is consistent with a particular axiom and a small
subset $\Sigma^*$ of $\Sigma$. To identify this subset of CI statements, we
examine the overlap between $\gamma$ and each CI statement in $\Sigma$,
selecting only those with non-empty overlaps. In this context, we define the
overlap between $\gamma$ and $\sigma$. Without loss of generality, let $\gamma$
be $\bm{X}\ci \bm{Y}\mid \bm{Z}$ and $\sigma$ be $\bm{X}'\ci \bm{Y}'\mid
\bm{Z}'$. Then, the overlap between $\gamma$ and $\sigma$ is defined as the
intersection: $U=(\bm{X}\cup \bm{Y}\cup \bm{Z})\cap (\bm{X}'\cup \bm{Y}' \cup
\bm{Z}')$. If $U$ is non-empty, we deem $\sigma$ and $\gamma$ have non-empty
overlaps.

We can break down the full constraint set into several smaller sets and resolve
them concurrently along with the full constraint. If any smaller set is
inconsistent, it infers the full constraint's inconsistency, allowing an
immediate result return without waiting for complete constraint resolution. If
not, we can securely return the outcome derived from the solution of the full
constraint. In essence, $\mathcal{O}_3$ concentrates on maximizing modern CPUs'
parallelism and interrupting constraint solving early when feasible.

\subsubsection{The Overall Decision Procedure}
\label{subsubsec:soundness}
Our decision procedure for the CIR problem combines three optimizations
(\S~\ref{subsec:ma}-\S~\ref{subsubsec:slice}): and an off-the-shelf SMT solver,
as illustrated in Figure~\ref{fig:overview}. By integrating the three
optimizations, the four-stage procedure enhances the efficiency and scalability
of the CIR solving process, making it suitable for real-world usages.

\parh{Soundness and Completeness.}~Although we make certain levels of over- or
under-approximations in our three optimizations, we only take the conclusive
results from these optimizations and resort to the SMT solver in the absence of
a conclusive result. Our encoding of the CIR problem is faithful and does not
make any approximations. \revise{The functionality of de facto SMT solvers is
guaranteed by the SMT-LIB standard~\cite{barrett2010smt} and their
implementations are well-tested~\cite{winterer2020validating}. Hence, it should
be accurate to assume that modern SMT solvers (e.g., Z3~\cite{de2008z3} adopted
in this research) will return a correct result.} Hence, the overall procedure is
sound and complete with respect to Pearl's axioms.

\section{Implementation of \tool}
\label{subsec:impl}

Based on the CIR problem, we implement two checkers: \edsan\ and \psan, as
depicted in \A~\ref{alg:cisan}. We integrate \tool\ with the standard PC
algorithm, one highly popular causal discovery algorithm. However, \tool\ is
independent of the causal discovery algorithm and can be integrated with any
causal discovery algorithm as long as it uses CI tests (see
\S~\ref{sec:discussion} on extension). We implement \tool\ in Python with
approximately 2,000 lines of code. Our implementation is optimized for
multi-core parallelism. We use Z3~\cite{de2008z3} as the SMT solver. Our
implementation can be found at~\cite{artifact}. Next, we describe two design
choices in our implementation. 

\parh{Handling Inconsistent Knowledge Bases in \psan.}~\revise{Unreliable CI
tests may introduce errors into the knowledge base, and these errors may render
the knowledge base inconsistent (i.e., $\textit{KB}\models \bot$). In the
context of \edsan, we can simply abort the causal discovery process. However, in
\psan, it is desirable to continue the causal discovery process at the best of
efforts even with an inconsistent knowledge base. However, an inconsistent
knowledge base may cause our solving pipeline to fail to determine the
consistency of an incoming CI statement.} To address this issue, we backtrack to
the last consistent knowledge base and continue solving to the best of our
ability. If inconsistencies occur frequently, after reaching a threshold (ten in
our evaluation; see \S~\ref{sec:simulation}), we revert to using standard CI
tests for all CI statement queries.

\parh{Handling Timeout.}~Time constraints could cause the SMT solver to yield an
\texttt{unknown} when trying to determine the satisfiability of a constraint. As
a common tactic in program analysis, \texttt{unknown} is interpreted as
``satisfiable.'' This tactic could cause false negatives, wherein \edsan\
overlooks some incorrect CI statements and \psan\ misses some pruning
opportunities. Despite these, \tool\ remains complete because it spots wrong or
excessive CI statements based solely on unsatisfiable outcomes.

\parh{Application Scope of \edsan.}~\revise{Certain errors in causal discovery,
which do not contradict Pearl's axioms, may be inherently undetectable. While
adhering to Pearl's axioms is crucial, it is not the sole determinant of
reliability. Holistically, these errors can arise from two factors --- the
partial nature of Pearl's axioms and the flexibility of the probability
distribution.

In the former case, the completeness of Pearl's axioms was traditionally assumed
correct, but Studeny~\cite{studeny} disproved this by showing that some CI statements
complying with Pearl's axioms are not entailed by any probability distribution.
However, we argue that these CI statements pose minimal practical concerns due
to their rarity. In the latter case, the versatility of the probability
distribution presents a critical limitation to the CIR problem. For example, in
a dual-variable system ($X,Y$), both $X\ci Y$ and $X\notci Y$ conform to Pearl's
axioms. In this scenario, \edsan\ fails to spot the error as these incorrect CI
statements could perfectly align with another valid probability distribution.
Analogous to this, the control-flow integrity (CFI) checker cannot detect all
control-flow hijacking attacks~\cite{carlini2015control, conti2015losing} when
these attacks follow a valid control-flow path. Nevertheless, we conjecture that,
in normal usages of \edsan (excluding adversarial use), such cases are not
likely to happen.

Therefore, we expect \edsan\ to effectively uncover the majority of errors in
practical scenarios. Since Pearl's axioms serve as a stringent integrity
constraint, erroneous CI statements, even undetectable immediately, likely
breach Pearl's axioms eventually, especially when $\Sigma$ is non-trivially
large. As will be shown in \S~\ref{subsec:rq1} and also in
\S~\ref{sm:detectability} of \sm, \edsan\ is highly sensitive in detecting
subtle errors in practice.}
\section{Evaluation}
\label{sec:simulation}

In this section, we study the effectiveness of our framework by answering the
following research questions:
\begin{enumerate}
  \item[\textbf{RQ1}] Does \edsan\ achieve high error detectability on
  erroneous CI statements?
  \item[\textbf{RQ2}] Does \psan\ effectively prune excessive CI tests in the
  privacy-critical setting?
  \item[\textbf{RQ3}] How effective are our optimizations?
\end{enumerate}

\subsection{Evaluation Setup}

\begin{table}[h]
  \vspace{-10pt}
  \centering
  \caption{Datasets.} 
  \vspace{-12pt}
  \resizebox{0.80\linewidth}{!}{
  \begin{tabular}{l|c|c|c|c}
    \hline
    \textbf{Dataset} & \textbf{\#Nodes} & \textbf{\#Edges} & \textbf{Max Degree}& \textbf{Max Indegree}\\
    \hline
    Earthquake  & 5 & 4 & 4 & 2 \\\hline
    Cancer  & 5 & 4 & 4 & 2 \\\hline
    Survey  & 6 & 6 & 4 & 2 \\\hline
    Sachs  & 11 & 17& 7  & 3  \\\hline
  \end{tabular}
   }
   \vspace{-5pt}
\label{tab:datasets}
\end{table}

We use four popular Bayesian networks from the bnlearn repository~\cite{bnlearn}
and list their details in \T~\ref{tab:datasets}. They come from diverse fields
and some of them even contain highly skewed data distributions which are very
challenging for CI tests (e.g., the ``Earthquake'' dataset). Overall, we clarify
that the evaluated datasets are practical and challenging, manifesting
comparable complexity to recent works with a similar
focus~\cite{zhalama2019asp,hyttinen2013discovering}. \revise{We illustrate the
complexity of CIR problem instances involved in the benchmark in
\S~\ref{sm:complexity} of \sm. Besides, we further
follow~\cite{xu2017differential} to evaluate \psan (in RQ2) with three synthetic
datasets ranging from five to 15 variables. The preparation procedure is
detailed in \S~\ref{sm:prep} of \sm.} In RQ1, we assess \edsan\ by first using a
perfect CI oracle that returns the ground-truth CI statement. Then, we randomly
inject errors with different scales into the CI statements and assess the
effectiveness of \edsan\ in error detection. In RQ2, we explore the use of
\psan\ in two imperfect CI test schemes in addition to the perfect CI oracle,
where the $\chi^2$ test is for normal causal discovery and Kendall's $\tau$
test~\cite{wang2020towards} is for differentially private causal discovery. To
handle inconsistent knowledge bases (noted in \S~\ref{subsec:impl}), we resort
to the standard CI tests when inconsistencies occur over ten times. All
experiments are conducted on a server with an AMD Threadripper 3970X CPU and 256
GB memory. 

\subsection{Effectiveness of \edsan}
\label{subsec:rq1}

In line with prior research~\cite{chen2016parametric}, we utilize a
collection of manually-crafted erroneous cases to assess whether \edsan\ can
detect them. Given a dataset from \T~\ref{tab:datasets}, we initially integrate
the causal discovery process with \edsan\ without injecting additional errors,
verifying that no false alarms arise from \edsan. Let $K$ represents the total
number of CI tests conducted on this dataset and $r\%$ an error injection rate.
We then randomly select $k = K \times r\%$ indices from $1, \ldots, k$ ($e_1,
\ldots, e_k$) and inject errors by flipping the outcome of the $e_i$-th CI test
for each $e_i \in \{e_1, \ldots, e_k\}$. It is important to note that \edsan\ is
not aware of specific error injection locations and validates errors after every
CI test. The process is repeated for $r \in \{1, \ldots, 10\}$, ensuring at
least one error per run.

\begin{figure}[t]
  \centering
  \includegraphics[width=0.7\linewidth]{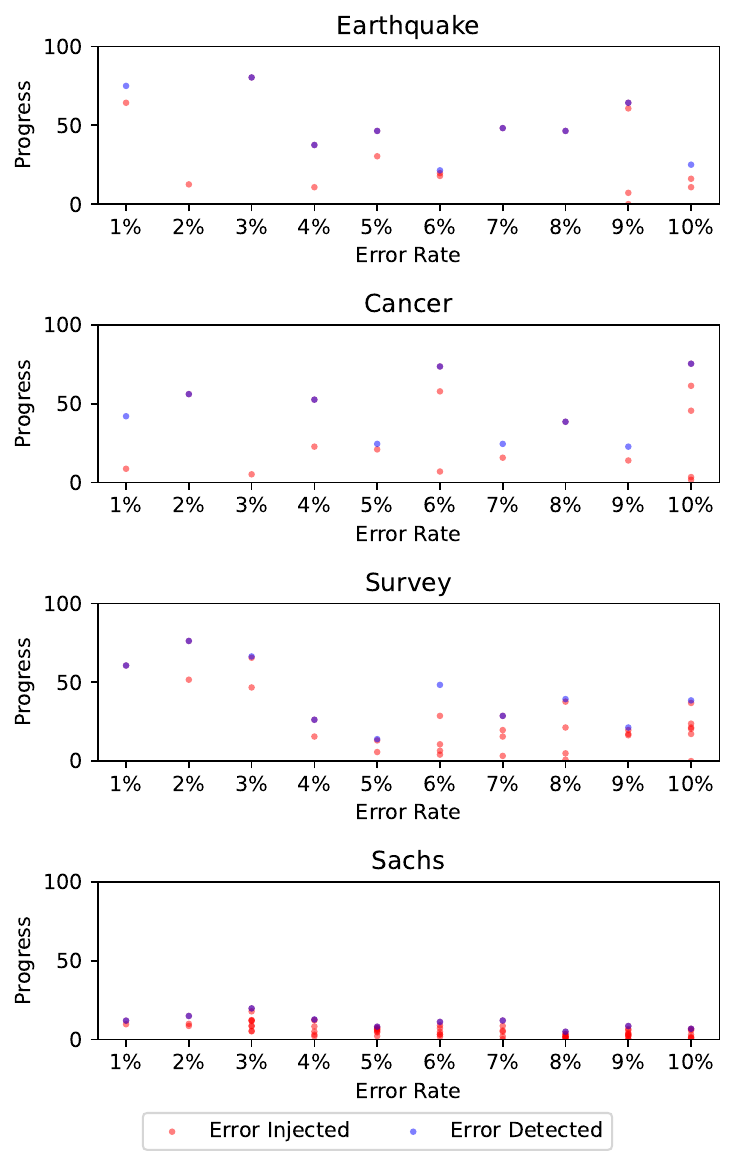}
  \vspace{-12pt}
  \caption{\edsan\ on four datasets.}
  \label{fig:rq1}
\end{figure}

\begin{table}[t]
  \vspace{-5pt}
  \centering
  \caption{Sensitivity of error detection.}
  \vspace{-12pt}
  \resizebox{0.7\linewidth}{!}{
  \begin{tabular}{l|c|c|c|c}
    \hline
    \textbf{Dataset} & Earthquake & Cancer & Survey & Sachs\\
    \hline
    \textbf{Sensitivity}      & 15.1\% & 25.0\% & 21.5\% & 8.0\%  \\\hline
  \end{tabular}
   }
\label{tab:rq1-sensitivity}
   \vspace{-5pt}
\end{table}

We present the error detection results in \F~\ref{fig:rq1}. Red dots represent
the normalized position (by the total number of CI tests) where errors are
injected, and blue dots indicate where \edsan\ detects them. Overlapping red and
blue dots indicate immediate error detection upon injection. We observe that
\edsan\ successfully detects errors in most cases (38 out of 40), excluding
``2\%'' of Earthquake and ``3\%'' of Cancer. Error detection sensitivity is also
quantified in Table~\ref{tab:rq1-sensitivity}, calculated by subtracting the
normalized progress of the first error detection from that of the error
injection. For instance, sensitivity in \F~\ref{fig:rq1} is determined by the
distance between the first red and blue dot in each plot. Overall, \edsan\
exhibits high sensitivity to errors, with six detected immediately and 25 within
15\% of the total CI tests.

\smallskip
\parh{Failure Case Analysis.}~We also inspected the two failure cases in
\F~\ref{fig:rq1}. In both cases, errors are injected into a marginal CI statement.
As aforementioned in \S~\ref{subsec:ma}, marginal CI statements themselves would
not violate Pearl's axioms. If there is no other high-order CI statements that
trigger the inconsistency, \edsan\ cannot detect the error by design.
Unfortunately, the injected errors do not violate any high-order CI statements.
Indeed, despite the error, we find that the learned causal skeleton is identical
to the ground truth. Hence, we presume these errors may be too subtle to be
detected.

\parh{Answer to RQ1:}~\textit{When benchmarked using common causal discovery
datasets with sufficient complexity, \edsan\ uncovers erroneous CI statements
with high accuracy and also high sensitivity.}

\subsection{Effectiveness of \psan}
\label{subsec:rq2}

In this evaluation, we employ the perfect CI oracle as well as two distinct CI
test schemes to evaluate the performance of \psan\ in reducing the number of CI
tests produced by the PC algorithm. The first CI test scheme is the
well-established $\chi^2$ test~\cite{pearson1900x}, a statistical hypothesis
test that assesses the independence of two categorical variables. This test is
widely used in causal discovery and often serves as the default CI test scheme
in many causal discovery tools. Hence, the result of the $\chi^2$ test provides
a baseline for assessing the effectiveness of \psan\ in realistic settings. The
second scheme is the differentially private Kendall's $\tau$
test~\cite{wang2020towards}, which is designed to protect the privacy of the
dataset by introducing controlled noise into the test statistics. This test is
particularly useful for evaluating the performance of \psan\ in
privacy-sensitive applications, where the traditional $\chi^2$ test may not be
suitable due to the disclosure risks associated with its non-private nature. By
incorporating these two CI test schemes, we aim to demonstrate \psan's
applicability in handling diverse CI test schemes.

\begin{table}[t]
  \vspace{-5pt}
  \centering
  \caption{Performance with the perfect CI Oracle.}
  \vspace{-10pt}
  \resizebox{0.75\linewidth}{!}{
  \begin{tabular}{l|c|c}
    \hline
    \textbf{Dataset} & \#CI Tests w/o \psan & \#CI Tests w/ \psan\\
    \hline
    Earthquake  & 56 & 22 (-60.7\%)  \\\hline
    Cancer      & 57 & 22 (-61.4\%) \\\hline
    Survey      & 115 & 51 (-55.7\%)   \\\hline
    Sachs       & 1258 & 818 (-35.0\%) \\\hline
  \end{tabular}
   }
   \vspace{-5pt}
\label{tab:perfect-ci}
\end{table}

\parh{\psan\ with Perfect CI Oracle.}~We present the results of our experiments
with the perfect CI oracle on four datasets in Table~\ref{tab:perfect-ci}. We
compared the number of CI tests generated by the PC algorithm with and without
\psan\ using a perfect CI oracle. The results are shown in
Table~\ref{tab:perfect-ci}, where we observe that \psan\ is highly successful in
decreasing the number of required CI tests across all datasets. Specifically,
for datasets with five or six variables, namely Earthquake, Cancer, and Survey,
we achieve a reduction ranging from 55.7\% to 60.7\%. Even on the challenging
Sachs dataset with 11 variables, where the UF to a CI statement can have up to
$2^{33}$ states, \psan\ demonstrates a significant reduction of 35.0\%.

As discussed in \S~\ref{subsubsec:soundness}, the CIR problem solving is sound
and complete, which means that \psan\ does not yield any erroneous CI
statements. Indeed, we observe that the learned causal skeleton is identical to
the ground-truth skeleton when using the perfect CI oracle, with no additional
errors introduced.

\begin{table}[t]
  \centering
  \caption{Effectiveness of \psan\ in terms of \#CI Test, Structural Hamming
  Distance (SHD) and privacy budget ($\epsilon$). ``ER-$n$'' denotes the
  synthetic dataset with $n$ variables and ``ER-15'' denotes the largest
  synthetic dataset with 15 variables.}
   \vspace{-8pt}
  \setlength{\tabcolsep}{1.5pt}
  \resizebox{\linewidth}{!}{
  {\color{revisecolor}\begin{tabular}{l|c|c|c|c|c|c|c}
    \hline
    \textbf{Metric} & Earthquake & Cancer & Survey & Sachs & ER-5 & ER-10 & ER-15 \\\hline
    \multicolumn{8}{c}{DP Kendall's $\tau$ Test (averaged on ten runs)}\\
    \hline
    \#CI Test    & 37.7$\to$16.2 & 16.8$\to$10.7& 45.9$\to$16.1 & 264.3$\to$156.2 & 13.3$\to$10.0 & 86.7$\to$67.0 & 190.0$\to$153.3  \\\hline
    SHD          & 2.7$\to$2.6   & 2.8$\to$2.6  & 2.5$\to$0.7   & 13.9$\to$13.0 & 1.7$\to$2.0        & 7.3$\to$7.0       & 20.3$\to$18.7 \\\hline
    $\epsilon$   & 54.6$\to$34.3 & 38.3$\to$32.1& 60.2$\to$40.8 & 277.9$\to$208.5& 35.2$\to$29.9     & 141.7$\to$118.8     & 280.6$\to$238.4 \\\hline
    \multicolumn{8}{c}{$\chi^2$ Test}\\\hline
    \#CI Test    & 56$\to$22 & 75$\to$36 & 69$\to$30 & 1277$\to$816& 15$\to$11 & 208$\to$136 & 610$\to$475 \\\hline
    SHD  & 0$\to$0 & 2$\to$2& 0$\to$0 & 1$\to$1& 0$\to$0 & 1$\to$1 & 4$\to$4  \\\hline
  \end{tabular}}
   }
   \vspace{-5pt}
\label{tab:imperfect-ci}
\end{table}

\parh{\psan\ with Imperfect CI Tests.}~\revise{The efficacy of \psan\ is 
validated on two imperfect CI test schemes: DP Kendall's $\tau$ and $\chi^2$
Tests. The results are in \T~\ref{tab:imperfect-ci}. SHD represents the ``edit
distance'' from the learned graph to the ground truth; a lower value is
preferable. Total privacy for executing DP Kendall's $\tau$ tests is denoted by
$\epsilon$; a lower value is preferable. Owing to the probabilistic character of
the DP-enforced algorithm, the DP Kendall's $\tau$ Test is repeated ten times
and the average outcome is reported. \psan\ efficiently reduces the necessary CI
tests count while maintaining competitive SHD accuracy across all data sets. On
average, \psan\ results in a reduction of 38.0\% and 41.2\% in CI tests using
DP Kendall's $\tau$ and $\chi^2$ Tests respectively, across all datasets.
Moreover, there is an observable slight increase in accuracy (lower SHD) under
the DP setting compared to Priv-PC~\cite{wang2020towards}. This is presumably
because \psan\ sidesteps the high number of CI tests that are typically
incorrect. In addition, \psan offers superior privacy protection, as indicated
by a substantially lower $\epsilon$.}

\parh{Answer to RQ2:}~\textit{\psan\ effectively prunes excessive CI tests in
both perfect and imperfect CI test schemes. It reduces CI tests by 35.0\% to
60.7\% with the perfect CI oracle and retains accuracy while providing better
privacy protection in imperfect CI tests.}

\subsection{Ablation Study}
\label{subsec:rq3}

In RQ3, we aim to evaluate the effectiveness of our proposed optimizations and
explore their impact on the overall performance. Recall that there are three
optimizations in \tool, namely $\mathcal{O}_1$ MA, $\mathcal{O}_2$ GR, and
$\mathcal{O}_3$ CD. In particular, $\mathcal{O}_1$ is used to prove consistency
while $\mathcal{O}_2$ and $\mathcal{O}_3$ are used to refute consistency.
\revise{Thus, given their different purposes, we employ \textit{distinct
evaluation strategies} for $\mathcal{O}_1$ and the other two optimizations. We
measure $\mathcal{O}_1$'s effectiveness through the entire lifecycle of the
causal discovery. On the other hand, $\mathcal{O}_2, \mathcal{O}_3$ are
particularly effective when dealing with inconsistent CI statements. In \edsan,
there is at most one inconsistent case (since we abort execution once an
inconsistency is detected). Hence, we measure the effectiveness of
$\mathcal{O}_2, \mathcal{O}_3$ on multiple instances of the CIR problem. Below,
we present the impact of them.}

\begin{figure}[t]
  \vspace{-3pt}
  \centering
  \includegraphics[width=0.9\linewidth]{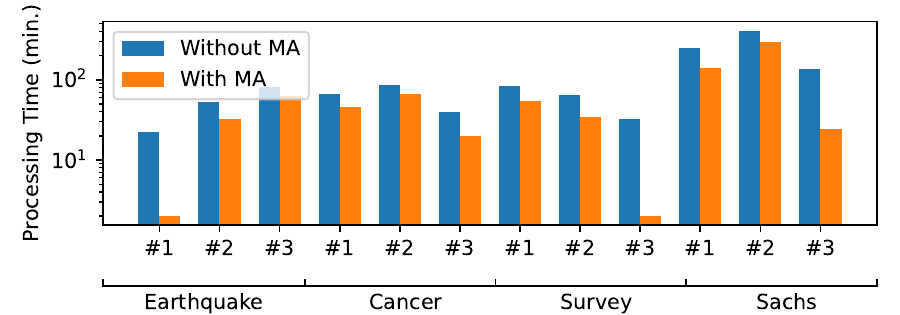}
  \vspace{-12pt}
  \caption{Comparison of processing time (each bar represents an individual run).}
  \vspace{-12pt}
  \label{fig:ma}
\end{figure}

\parh{Effectiveness of $\mathcal{O}_1$ MA.}~The effectiveness of $\mathcal{O}_1$
is first evaluated through a comparison of processing times between \edsan\ with
and without $\mathcal{O}_1$. Our experiment involves injecting 5\% errors into
CI statements. The random seed remains fixed per comparison group. Each dataset
undergoes three comparative groupings. As shown in \F~\ref{fig:ma},
$\mathcal{O}_1$ notably reduces the time. Specifically, \tool\ immediately flags
the error in the initial non-marginal CI statements in the first case of
Earthquake and the third case of Survey. As a result, in such situations,
$\mathcal{O}_1$ makes \tool\ almost effortless. On average, $\mathcal{O}_1$
achieves a speedup of 48.9\% ($\pm$ 25.7\%) across all datasets, with a maximum
reduction of 93.8\%. Our interpretation is that $\mathcal{O}_1$ effectively
accelerates \tool\ without compromising accuracy.

\parh{Effectiveness of $\mathcal{O}_2$ GR and $\mathcal{O}_3$ CD.}~We then
evaluate the effectiveness of $\mathcal{O}_2$ and $\mathcal{O}_3$.  To fully
unleash the power of $\mathcal{O}_2$ and $\mathcal{O}_3$, we use the Sachs
dataset, which is the largest dataset in our experiments. On the remaining
datasets, we observe similar performance boosts from $\mathcal{O}_2$ and
$\mathcal{O}_3$ (see \S~\ref{sm:extended-experiments-rq3} of \sm). In
particular, we collect a total of 1,258 CI statements from the error-free
execution on the Sachs dataset. Then, we randomly flip some CI statements (as
well as their symmetric counterparts) to create a knowledge base such that each
knowledge base contains roughly 5\% erroneous CI statements. Hence, we obtain
one instance of the CIR problem with erroneous CI statements that may be
inconsistent with the Pearl's axioms. We check whether the inconsistent
knowledge bases can be refuted by $\mathcal{O}_2$ or $\mathcal{O}_3$,
respectively. We also measure the processing time. We repeat the experiments 100
times and report the average running time and the overall refutability. 

\begin{table}[t]
  % \vspace{-5pt}
  \centering
\caption{Effectiveness of $\mathcal{O}_2$ GR and $\mathcal{O}_3$ CD. ``avg.'',
``mdn.'' and ``max.'' denotes the mean, median and maximum.}
\label{tab:rq3-effectiveness}
\vspace{-10pt}
\resizebox{0.8\linewidth}{!}{
{\color{revisecolor}\begin{tabular}{l|c|c}
  \hline
    & Refutability & Processing time (avg. / mdn. / max.) \\
  \hline
  $\mathcal{O}_2$ & 91/100 & 0.6 / {0.6} / {0.7}\\
  \hline
  $\mathcal{O}_3$ w/o Full & 53/100 & 14.3 / {10.9} / {25.8} \\
  \hline
  $\mathcal{O}_3$ & 100/100 & 25.6 / {5.3} / {58.2} \\
  \hline
  $\mathcal{O}_2$ + $\mathcal{O}_3$ & 100/100 & 2.4 / {0.6} / {53.8} \\
  \hline
  \end{tabular}}
  }
  \vspace{-12pt}
\end{table}

\revise{As shown in \T~\ref{tab:rq3-effectiveness}, in a series of 100
experiments, $\mathcal{O}_2$ has an efficacy of refuting 91\% of the
inconsistent knowledge bases, and $\mathcal{O}_3$ --- when combined with full
constraint solving --- manages to refute 100\% of the inconsistencies, while the
decomposed constraints alone can only address 53\% of the errors. When both
$\mathcal{O}_2$ and $\mathcal{O}_3$ are applied together, they detect all
errors, enhancing their collective effectiveness. We notice that the median
processing time of $\mathcal{O}_2$ is significantly lower (0.6 seconds) as
compared to that of using decomposed constraints alone, which requires 10.9
seconds, and extends to 5.3 seconds when used alongside the full constraint,
marking $\mathcal{O}_2$ as both efficient and highly sensitive towards
inconsistent knowledge bases. Besides, we observe notable variations in the
mean, median, and maximum processing time for $\mathcal{O}_2$ + $\mathcal{O}_3$.
We interpret the result as follows. The processing time primarily clusters
around two scenarios; the majority approximating 0.6 seconds (case 1), with the
remainder around 53.8 seconds (case 2). The first case corresponds to situations
where $\mathcal{O}_2$, a lightweight scheme, swiftly refutes consistency. The
latter case occurs when an SMT solver is involved and requires more time to
resolve the CIR problem.

On the other hand, $\mathcal{O}_3$ provides a comprehensive overview of the CIR
problem and detects all errors, serving as an effective complement to
$\mathcal{O}_2$ by refuting additional inconsistent knowledge bases. From
another perspective, $\mathcal{O}_2$ significantly reduces the costs associated
with $\mathcal{O}_3$ by detecting numerous errors early on. Simultaneous
adoption of $\mathcal{O}_2$ and $\mathcal{O}_3$ trims down the processing time
to 2.3 seconds, yielding a solution that is more accurate than using
$\mathcal{O}_2$ alone, and remarkably faster than just utilizing
$\mathcal{O}_3$.}

\parh{Answer to RQ3:}~\textit{All three optimizations significantly contribute
to the performance of \tool. Specifically, $\mathcal{O}_1$ reduces the
processing time by a significant margin, while $\mathcal{O}_2$ and
$\mathcal{O}_3$ are effective and efficient in detecting and refuting
inconsistent knowledge bases. The synergistic effect of these optimizations
makes \tool\ practical.}
\subsection{Discussion}
\label{sec:discussion}

\parh{Potential Impact in Software Engineering.}~\revise{We articulate the
potential impact of our work in the SE community from two perspectives.
\ding{202} {\it Causal Inference in SE:} Causal inference is gaining traction as
an important methodology within SE~\cite{fariha2020causality,
dubslaff2022causality, johnson2020causal, hsiao2017asyncclock, ji2022perfce,
siebert2023applications, ji2022perfce, ji2023causality, sun2022causality,
zhang2022adaptive, ji2023cc}. The prosperous use of causal inference algorithms
makes \tool\ an important contribution that uniquely addresses both privacy and
reliability concerns. This positions \tool\ as a useful tool in the
fast-evolving SE landscape. \ding{203} {\it Methodological Integration:} \tool\
delivers a novel adoption of classical SE techniques, e.g., formal verification
and constraint solving. This synergy, to a degree, enables \tool\ to expand the
existing formal method realm into a novel and crucial domain --- causal
discovery.}

\parh{Scalability.}~Automated reasoning is often costly, and the scalability of
our framework is on par with that of other SMT/SAT-based causal discovery
algorithms~\cite{zhalama2019asp,hyttinen2013discovering} \revise{(e.g., causal
graph with roughly a dozen of nodes)}. Causal discovery algorithms often
struggle with high-dimensional data, which is why \edsan\ focuses on datasets
with moderate complexity; de facto causal discovery algorithms are anticipated
to behave reliably and perform well over these datasets. \revise{In terms of
\psan, existing constraint-based privacy-preserving causal discovery algorithms,
such as EM-PC~\cite{xu2017differential}, share a similar scalability regime.
However, Priv-PC~\cite{wang2020towards} shows a better scalability which handles
larger datasets like ALARM and INSURANCE~\cite{bnlearn}. Notably, \tool\ has a
unique role compared to these algorithms, potentially enhancing their
privacy-preserving abilities, evidenced by \psan's improvement on Priv-PC
(\S~\ref{subsec:rq2}).} Looking ahead, we plan to boost \tool\ with advanced SMT
solvers like CVC5~\cite{barbosa2022cvc5}.

\parh{Extension to Other Causal Discovery Algorithms.}~Since \tool\ does not
rely on any execution information about the underlying algorithms, it is
agnostic to specific causal discovery implementations. Hence, our framework can
be smoothly extended to other causal discovery algorithms by replacing the
employed standard CI test algorithm with the one hardened by \tool. For
instance, we envision that \tool\ can be applied to enhance the federated causal
discovery algorithm~\cite{wang2023towards} to reduce privacy leakage. However,
\tool is not applicable to enhance causal discovery algorithms that do not use
CI tests (e.g., functional causal models~\cite{peters2017elements,
ma2022noleaks} or score-based methods~\cite{tsamardinos2006max}) or leverage
quantitative information beyond qualitative CI relations (e.g.,
p-values~\cite{ma2022ml4s, ding2020reliable}).

\parh{Threats to Validity.}~Our framework is implemented using standard tool
chains and the experiments are conducted with widely-used datasets and CI test
algorithms. Nonetheless, potential threats to the validity of our results
persist. Firstly, our framework's performance may be contingent on the choice of
CI test algorithms and datasets. To alleviate this risk, we have tested our
framework with multiple CI test algorithms and datasets. Secondly, our
framework's scalability may face limitations when employed on large-scale
datasets or complex causal structures. To address this, we propose several
optimizations to improve efficiency. Thirdly, the generalizability of our
results may be influenced by the specific implementation choices within our
framework. To mitigate this threat, we conducted experiments on several datasets
and CI test algorithms repetitively to ensure the robustness of our results.
Besides, using random error injection in \S~\ref{subsec:rq1} may not perfectly
simulate real-world cases. In this regard, our findings' applicability might be
limited. Integrating \tool\ with causal learning tools~\cite{causallearn} would
be a beneficial way to confirm the validity. As mentioned in
\S~\ref{subsec:impl}, it is theoretically conceivable that multiple incorrect
facts could weaken the effectiveness of \edsan. While we deem this scenario rare
due to the tight integrity constraints by Pearl's axioms and empirically show
its unlikelihood in practice (\S~\ref{subsec:rq2} and \S~\ref{sm:detectability}
of \sm), we acknowledge it as a threat to validity. Statistical errors could
cause \psan to malfunction because of inconsistent knowledge bases or faulty
reasoning (i.e., deriving an incorrect CI from another wrong CI). We address
inconsistencies in the knowledge bases by employing a backtracking mechanism to
restore previous states (\S~\ref{subsec:impl}). We demonstrate empirically that
\psan does not compromise causal discovery accuracy, proving its resilience to
statistical errors (\S~\ref{subsec:rq2}).

\section{Related Work}

We have reviewed the related works in causal discovery in \S~\ref{sec:prel}. In
this section, we discuss the related work in formal methods for CI implication
and causal inference in software engineering.

\parh{CI Implication.}~Some studies aim to infer CI statements through
implications. Instead of using Pearl's axioms, they attempt to characterize
specific classes of distribution and infer CI statements using algebraic or
geometric reasoning methods~\cite{bouckaert2007racing, bouckaert2010efficient,
tanaka2015linear, niepert2013conditional}. It is worth noting that these
approaches are not applicable in our context. Firstly, they cannot reason over
conditional dependence --- a prevalent and essential inference case in our
scenario. Secondly, they often exhibit limited scalability and narrow
applicability (e.g., the racing-based method~\cite{bouckaert2007racing} can only
handle five variables).

\parh{Causality in Software Engineering.}~Causality has become an increasingly
prevalent technique in software engineering~\cite{fariha2020causality,
dubslaff2022causality, johnson2020causal, hsiao2017asyncclock, ji2022perfce,
siebert2023applications, ji2022perfce, ji2023causality}. These tools typically
concentrate on examining the effects of program inputs or configurations on
software behavior, such as execution time or crashes. Moreover, due to its
inherent interpretability, causality analysis is extensively utilized in the
testing and repair of machine learning models~\cite{sun2022causality,
zhang2022adaptive, ji2023cc}.

\section{{Conclusion}}
\label{sec:con}

We have presented \tool, a novel runtime verification tool for causal discovery
algorithms that addresses both reliability and privacy concerns in
\textit{reliability-critical} and \textit{privacy-critical} usage scenarios in a
unified approach. Experiments on various synthetic and real-world datasets show
that \tool\ enhances both reliability-critical and privacy-critical scenarios
with a moderate overhead. \revise{We envision our work can inspire more research
on the extension of SE methodology to the domain of causal inference.}

\begin{acks}

The authors would like to thank the anonymous reviewers for their valuable
comments. The authors would also like to thank Rui Ding, Haoyue Dai, Yujia Zheng
and Joseph Ramsey for their helpful discussions. This work is supported in part
by National Key R\&D Program of China under Grant No. 2022YFB4501903, the HKUST
30 for 30 research initiative scheme under the the contract Z1283, National
Natural Science Foundation of China under Grant No. 62302434, and the Qizhen
Scholar Foundation of Zhejiang University.

\end{acks}

\balance
\bibliographystyle{ACM-Reference-Format}
\bibliography{main}


\clearpage
\onecolumn
\appendix

\section{{Preliminaries}}
\label{sm:prel}
\revise{\parh{d-separation.}~In a DAG, two sets of nodes $\bm{X}$ and $\bm{Y}$
are said to be d-separated by $\bm{Z}$ if and only if there is no active
undirected path between $\bm{X}$ and $\bm{Y}$ given $\bm{Z}$. An undirected path
between $X\in \bm{X}$ and $Y\in\bm{Y}$ is active relative to $\bm{Z}$ if (i)
every non-collider on $U$ is not a member of $\bm{Z}$ and (ii) every collider on
$U$ has a descendant in $\bm{Z}$. Here, $U$ is said to be a collider on the path
if the edges connected $U$ form the structure of $\to U\leftarrow$. Intuitively,
the $d$-separation relation can be seen as a graphical analogy of CI in data. In
other words, all paths being deactivated by $\bm{Z}$ indicates that the causal
relation between $\bm{X}$ and $\bm{Y}$ is blocked by $\bm{Z}$.}

\revise{\parh{Faithful Bayesian Network.}~A Bayesian network is said to be
faithful if and only if it satisfies the following two conditions: (i) the
Markov condition and (ii) the faithfulness condition. The Markov condition
states that every node is independent of its non-descendants given its parents.
The faithfulness condition states that every CI in the distribution is entailed
by the DAG. In other words, the faithfulness condition states that the DAG is
minimal in the sense that removing any edge from the DAG will result in a
distribution that does not entail the CI.}

\section{Encoding of Pearl's Axioms}
\label{sm:enc}
\noindent\textbf{Symmetry.}
\begin{equation}
    \begin{aligned}
        \forall &BV_{\bm{X}},BV_{\bm{Y}},BV_{\bm{Z}}\in \{0,1\}^{|\bm{V}|}\\
        & \texttt{CI}(BV_{\bm{X}}, BV_{\bm{Y}}, BV_{\bm{Z}}) = \texttt{CI}(BV_{\bm{Y}}, BV_{\bm{X}}, BV_{\bm{Z}})\\
    \end{aligned}
\end{equation}
\textbf{Decomposition.}
\begin{equation}
    \begin{aligned}
        \forall &BV_{\bm{X}},BV_{\bm{Y}},BV_{\bm{Z}},BV_{\bm{W}}\in \{0,1\}^{|\bm{V}|}\\
        & \texttt{CI}(BV_{\bm{X}}, BV_{\bm{Y}} | BV_{\bm{W}}, BV_{\bm{Z}})=\texttt{01}\implies \texttt{CI}( BV_{\bm{X}}, BV_{\bm{Y}}, BV_{\bm{Z}})=\texttt{01}\\
        & \texttt{CI}( BV_{\bm{X}}, BV_{\bm{Y}} | BV_{\bm{W}}, BV_{\bm{Z}})=\texttt{01}\implies \texttt{CI}(BV_{\bm{X}}, BV_{\bm{W}}, BV_{\bm{Z}})=\texttt{01}
    \end{aligned}
\end{equation}
\textbf{Weak Union.}
\begin{equation}
    \begin{aligned}
        \forall &BV_{\bm{X}},BV_{\bm{Y}},BV_{\bm{Z}},BV_{\bm{W}}\in \{0,1\}^{|\bm{V}|}\\
        & \texttt{CI}(BV_{\bm{X}}, BV_{\bm{Y}} | BV_{\bm{W}}, BV_{\bm{Z}})=\texttt{01}\implies \texttt{CI}( BV_{\bm{X}}, BV_{\bm{Y}}, BV_{\bm{Z}} | BV_{\bm{W}})=\texttt{01}
    \end{aligned}
\end{equation}
\textbf{Contraction.}
\begin{equation}
    \begin{aligned}
        \forall &BV_{\bm{X}},BV_{\bm{Y}},BV_{\bm{Z}},BV_{\bm{W}}\in \{0,1\}^{|\bm{V}|}\\
        &  \texttt{CI}(BV_{\bm{X}}, BV_{\bm{Y}}, BV_{\bm{Z}})=\texttt{01}\land 
        \texttt{CI}(BV_{\bm{X}}, BV_{\bm{W}}, BV_{\bm{Y}} | BV_{\bm{Z}})=\texttt{01}\implies \texttt{CI}( BV_{\bm{X}}, BV_{\bm{Y}} | BV_{\bm{W}}, BV_{\bm{Z}})=\texttt{01}
    \end{aligned}
\end{equation}
\textbf{Intersection.}
\begin{equation}
    \begin{aligned}
        \forall &BV_{\bm{X}},BV_{\bm{Y}},BV_{\bm{Z}},BV_{\bm{W}}\in \{0,1\}^{|\bm{V}|}\\
        &  \texttt{CI}(BV_{\bm{X}}, BV_{\bm{Y}}, BV_{\bm{Z}}|BV_{\bm{W}})=\texttt{01}\land 
        \texttt{CI}(BV_{\bm{X}}, BV_{\bm{W}}, BV_{\bm{Y}} | BV_{\bm{Z}})=\texttt{01}\implies \texttt{CI}( BV_{\bm{X}}, BV_{\bm{Y}} | BV_{\bm{W}}, BV_{\bm{Z}})=\texttt{01}
    \end{aligned}
\end{equation}
\noindent\textbf{Composition.}
\begin{equation}
    \begin{aligned}
        \forall &BV_{\bm{X}},BV_{\bm{Y}},BV_{\bm{Z}},BV_{\bm{W}}\in \{0,1\}^{|\bm{V}|}\\
        &  \texttt{CI}(BV_{\bm{X}}, BV_{\bm{Y}}, BV_{\bm{Z}})=\texttt{01}\land 
        \texttt{CI}(BV_{\bm{X}}, BV_{\bm{W}}, BV_{\bm{Z}})=\texttt{01}\implies \texttt{CI}( BV_{\bm{X}}, BV_{\bm{Y}} | BV_{\bm{W}}, BV_{\bm{Z}})=\texttt{01}
    \end{aligned}
\end{equation}
\textbf{Weak Transitivity.}
\begin{equation}
    \begin{aligned}
        \forall &BV_{\bm{X}},BV_{\bm{Y}},BV_{\bm{Z}},BV_{\bm{W}},BV_{U}\in \{0,1\}^{|\bm{V}|}\\
        &  |BV_{U}| = 1\land
        \texttt{CI}(BV_{\bm{X}}, BV_{\bm{Y}}, BV_{\bm{Z}})=\texttt{01}\land 
        \texttt{CI}(BV_{\bm{X}}, BV_{\bm{W}}, BV_{\bm{Z}}|BV_{U})=\texttt{01}\implies \texttt{CI}(BV_{U}, BV_{\bm{Y}}, BV_{\bm{Z}})=\texttt{01}\\
        &  |BV_{U}| = 1\land
        \texttt{CI}(BV_{\bm{X}}, BV_{\bm{Y}}, BV_{\bm{Z}})=\texttt{01}\land 
        \texttt{CI}(BV_{\bm{X}}, BV_{\bm{W}}, BV_{\bm{Z}}|BV_{U})=\texttt{01}\implies \texttt{CI}( BV_{\bm{X}}, BV_{U},BV_{\bm{Z}})=\texttt{01}
    \end{aligned}
\end{equation}
\textbf{Chordality.}
\begin{equation}
    \begin{aligned}
        \forall &BV_{{X}},BV_{{Y}},BV_{{Z}},BV_{{W}}\in \{0,1\}^{|\bm{V}|}\\
        &  |BV_{X}| = 1\land|BV_{Y}| = 1\land|BV_{Z}| = 1\land|BV_{W}| = 1\land
        \texttt{CI}(BV_{{X}}, BV_{{Y}}, BV_{{Z}}|BV_{W})=\texttt{01}\land
        \texttt{CI}(BV_{Z}, BV_{{W}}, BV_{{X}}|BV_{Y})=\texttt{01}\\
        &\implies \texttt{CI}(BV_{X}, BV_{{Y}}, BV_{{Z}})=\texttt{01}\\
        &  |BV_{X}| = 1\land|BV_{Y}| = 1\land|BV_{Z}| = 1\land|BV_{W}| = 1\land
        \texttt{CI}(BV_{{X}}, BV_{{Y}}, BV_{{Z}}|BV_{W})=\texttt{01}\land
        \texttt{CI}(BV_{Z}, BV_{{W}}, BV_{{X}}|BV_{Y})=\texttt{01}\\
        &\implies \texttt{CI}(BV_{X}, BV_{{Y}}, BV_{{W}})=\texttt{01}
    \end{aligned}
\end{equation}

\section{{PC Algorithm}}
\label{sm:pc}
{\color{revisecolor}\begin{algorithm}[H]
\caption{PC Algorithm}
\label{alg:csl}
\SetKwFunction{Skl}{Skeleton-Learning}
\SetKwFunction{PC}{PC}
\SetKwFunction{FCI}{FCI}
\SetKwProg{Fn}{Function}{:}{}
\Fn{\Skl{$X_1,\cdots,X_d$}}{
    initialize a complete undirected graph $Q$ with $d$ nodes\;
    $n\leftarrow 0$\;
    
    \Repeat{for each ordered pair of adjacent nodes $X,Y$,
    $\textit{Neighbor}(X)\setminus \{Y\}$ has less than $n$ neighbors}{
    
        \Repeat{all ordered pairs of adjacent nodes $X$ and $Y$ such that
        $\textit{Neighbor}(X)\setminus \{Y\}$ has cardinality $\geq n$ and all
        subsets $\bm{S}$ in $\textit{Neighbor}(X)\setminus \{Y\}$ have been tested
        for d-separation}{
        select an ordered pair of adjacent nodes $X,Y$ such that
        $\textit{Neighbor}(X)\setminus \{Y\}$ has cardinality $\geq n$, and a subset
        $\bm{S}\subseteq \textit{Neighbor}(X)\setminus \{Y\}$ of cardinality $n$,
        and, if $X \Perp Y \mid S$, delete the edge between $X$ and $Y$ from $Q$,
        and record $\bm{S}$ in $\textit{Sepset}(X,Y)$ and $\textit{Sepset}(Y,X)$;
        }
        $n\leftarrow n+1$\; 
    }

    \KwRet{$Q$}
}

\Fn{\PC{$X_1,\cdots,X_d$}}{
    \tcc{skeleton learning}
    $G\leftarrow \Skl(X_1,\cdots,X_d)$\;
    \tcc{orientation}
    \ForEach{unshielded triple $(X-Z-Y)$ in $G$}{
        \uIf{$\forall X\ci Y\mid G, Z\notin G$}{
            orient $(X-Z-Y)$ as $(X\to Z\leftarrow Y)$;
        }
    }
    \tcc{apply Meek's rule~\cite{meek1995causal} to orient the remaining edges}
    \Repeat{no more edgs can be oriented}{
        \lForEach{$(X\to Z-Y)$}{orient as $(X\to Z\leftarrow Y)$}        
        \lForEach{$(X\to Z\to Y)$ and $(X-Y)$}{orient as $(X\to Y)$}        
        \lForEach{$(X-Y)$, $(X-Z)$, $(X-W)$, $(Z\to Y)$, $(W\to Y)$ and $Z,W$ is nonadjacent}{orient as $(X\to Y)$}        
    }
    \KwRet{$G$}
}
\end{algorithm}}

\revise{We present the workflow of the PC algorithm~\cite{spirtes2000causation}
in \A~\ref{alg:csl}. In the first step (lines 1--10; line 12), edge adjacency is
confirmed if there is no conditional independence between two variables (line
6). In the second step (lines 13--21), a set of orientation rules are applied
based on conditional independence and graphical criteria.}

\section{{Preparation of Synthetic Datasets}}
\label{sm:prep}
\revise{Following the setup in EM-PC~\cite{xu2017differential}, we sample random
graphs from the Erdős–Rényi graphical model~\cite{erdHos1960evolution} with
different node size. Then, we sample the corresponding joint distribution from
the Dirichlet-multinomial distribution (i.e., the family distributions of our
probability distributions). We use the default parameters from
EM-PC~\cite{xu2017differential} to generate the synthetic datasets.}

\section{{Complexity of Benchmark Problems}}
\label{sm:complexity}
\begin{table}[h]
  \centering
  \caption{\revise{Statistics of SMT Problems (avg. on ten runs).
  \texttt{\#quant-instantiations} indicates the number of instantiated
  quantifiers.}} \resizebox{0.5\linewidth}{!}{
  {\color{revisecolor}\begin{tabular}{l|c|c|c}
    \hline
    \textbf{Dataset} & \textbf{Memory Usage} (\texttt{MB}) & \textbf{Solving Time} (sec.) & \textbf{\texttt{\#quant-instantiations}}\\
    \hline
    Earthquake  & 20 & 0.3 & 192.0\\\hline
    Cancer  & 21 & 0.9 & 257.9 \\\hline
    Survey   & 22 & 0.8 & 337.2 \\\hline
    Sachs  & 96 & 33.9 & 2728.4 \\\hline
  \end{tabular}}
   }
\label{tab:smt-stat}
\end{table}

\revise{\parh{Number of SMT Constraints.}~Our encoding scheme posits a linear
relationship between SMT constraints and CI statements. Recall that Pearl's
axioms are enforced into a constant number of first-order constraints and each
CI statement is encoded into a propositional equality constraint on the
uninterpreted function $\texttt{CI}$. Therefore, with $n$ CI statements, we have
$n$ $\texttt{CI}$ equality constraints alongside $8$ first-order constraints,
totalling $n+8$ SMT Constraints. Accordingly, the max number of SMT constraints
per dataset in \T~\ref{tab:datasets} are 64, 65, 123 and 1266,
respectively. We consider such constraint sizes as moderate and can be smoothly handled by our technique.}

\revise{\parh{Statistics of SMT Problem Instances.}~In addition to the number of
SMT constraints, we present additional key metrics for each dataset's largest
SMT problem instance (the same instance in \S~\ref{subsec:rq3}) in
\T~\ref{tab:smt-stat}. The sizes of these problems are generally considered
moderate. On the other hand, for datasets like Earthquake, Cancer, and Survey,
the resolution time does not correlate highly with the size of CI statements.
This might be due to their relatively smaller sizes, making them more responsive
to external factors such as the particular optimizations employed by the SMT
solver.}

\section{{Detectability of \edsan}}
\label{sm:detectability}
\begin{figure}[ht]
    \centering
    \includegraphics[width=0.55\linewidth]{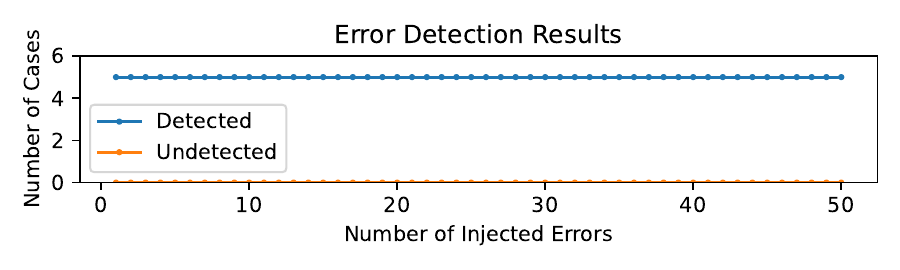}
    \caption{\revise{Detectability of \edsan on the Survey dataset under
    different error numbers. The best detectability is achieved when the
    ``Number of Cases'' for ``Detected'' is 5. That is, \edsan flags all
    erroneous cases.}}
    \label{fig:ed-san-survey-detectability}
  \end{figure}

\begin{figure}[ht]
  \centering
  \includegraphics[width=0.55\linewidth]{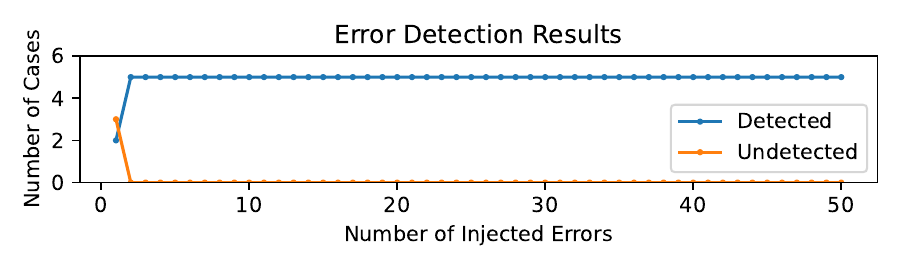}
  \caption{\revise{Detectability of \edsan on the Sachs dataset under different
  error numbers. The best detectability is achieved when the ``Number of Cases''
  for ``Detected'' is 5. That is, \edsan flags all erroneous cases.}}
  \label{fig:ed-san-sachs-detectability}
\end{figure}

\revise{We study the detectability of \edsan on the Survey and Sachs datasets
(representing two largest datasets in our experiments from the bnlearn repository).
In particular, we randomly inject one to fifty errors into the 1,258 CI
statements. We then run \edsan on the corrupted CI statements and record the
number of errors detected. We repeat this process five times with different
random seeds. The results are presented in
\F~\ref{fig:ed-san-survey-detectability} and
\F~\ref{fig:ed-san-sachs-detectability}. We observe that \edsan can
\textit{detect nearly all errors}. Notably, detection failures occur only in
three instances on the Sachs dataset (the largest dataset), when the error count
is one (corresponding to an error rate of $0.08\%$). Considering these
encouraging results, it is believed that \edsan offers highly practical
detectability, even for very subtle errors. }

\section{{End-to-end Effectiveness of $\mathcal{O}_2, \mathcal{O}_3$}}
\label{sm:e2e-psan}
\begin{table}[h]
\centering
\caption{\revise{End-to-end processing time of $\mathcal{O}_2$ GR and
$\mathcal{O}_3$ CD under when using \psan.}}
\label{tab:e2e-effectiveness-opt-psan}
\resizebox{0.4\linewidth}{!}{
{\color{revisecolor}\begin{tabular}{l|c|c|c|c}
\hline
 & Earthquake & Cancer & Survey & Sachs \\
\hline
$\mathcal{O}_1$ & 0:22:53 & 0:22:47 & 0:50:16 & 10:21:42 \\
\hline
$\mathcal{O}_1$ + $\mathcal{O}_2$ & 0:21:52 & 0:21:49 & 0:46:14 & 10:16:00 \\
\hline
$\mathcal{O}_1$ + $\mathcal{O}_3$ & 0:09:38 & 0:09:51 & 0:27:29 & 7:15:01 \\
\hline
Full Optimization & 0:09:24 & 0:07:35 & 0:25:21 & 7:03:15 \\
\hline
\end{tabular}}
}
\end{table}

\revise{We report the end-to-end processing time of \psan under different
optimization levels in \T~\ref{tab:e2e-effectiveness-opt-psan}. Since the
process of \psan does not involve any randomness, we only run each experiment
once. Overall, we view each optimization as effective, as it reasonably reduces the
processing time. In particular, $\mathcal{O}_3$ proves to be the most impactful
optimization by reducing the average processing time by 47.5\%. The effectiveness of
$\mathcal{O}_3$ is attributed to its lightweight constraint-solving scheme that
allows for quick disproval of the global consistency in CI statements and an
``early stop'' to the overall constraint-solving procedure. On the other hand,
we observe $\mathcal{O}_2$ is less effective than $\mathcal{O}_3$. Despite this,
$\mathcal{O}_2$ still contributes to an encouraging 4.4\% average increase in
speed. Overall, the full optimization scheme (i.e., $\mathcal{O}_1 +
\mathcal{O}_2 + \mathcal{O}_3$) delivers 51.8\% average speedup over the
baseline ($\mathcal{O}_1$ only).}

\section{Extended Experiments for RQ3}
\label{sm:extended-experiments-rq3}

\begin{table}[h]
  \centering
\caption{Effectiveness of $\mathcal{O}_2$ GR and $\mathcal{O}_3$ CD on Earthquake.}
\label{tab:rq3-effectiveness-earthquake}
\resizebox{0.35\linewidth}{!}{
\begin{tabular}{l|c|c}
\hline
    & Refutability & Avg. processing time (sec.) \\
\hline
$\mathcal{O}_2$ & 63/100 & 0.01 \\
\hline
$\mathcal{O}_3$ & 100/100 & 0.9 \\
\hline
$\mathcal{O}_2$ + $\mathcal{O}_3$ & 100/100 & 0.1 \\
\hline
\end{tabular}
}
\end{table}

\begin{table}[h]
  \centering
\caption{Effectiveness of $\mathcal{O}_2$ GR and $\mathcal{O}_3$ CD on Cancer.}
\label{tab:rq3-effectiveness-cancer}
\resizebox{0.35\linewidth}{!}{
\begin{tabular}{l|c|c}
\hline
    & Refutability & Avg. processing time (sec.) \\
\hline
$\mathcal{O}_2$ & 63/100 & 0.01 \\
\hline
$\mathcal{O}_3$ & 100/100 & 0.7 \\
\hline
$\mathcal{O}_2$ + $\mathcal{O}_3$ & 100/100 & 0.3 \\
\hline
\end{tabular}
}
\end{table}

\begin{table}[h]
  \centering
\caption{Effectiveness of $\mathcal{O}_2$ GR and $\mathcal{O}_3$ CD on Survey.}
\label{tab:rq3-effectiveness-survey}
\resizebox{0.35\linewidth}{!}{
\begin{tabular}{l|c|c}
\hline
    & Refutability & Avg. processing time (sec.) \\
\hline
$\mathcal{O}_2$ & 93/100 & 0.03 \\
\hline
$\mathcal{O}_3$ & 100/100 & 0.9 \\
\hline
$\mathcal{O}_2$ + $\mathcal{O}_3$ & 100/100 & 0.1 \\
\hline
\end{tabular}
}
\end{table}

\end{document}